\documentclass[aps,pra,superscriptaddress,twocolumn,a4paper,reprint,noeprint]{revtex4-1}
\usepackage[utf8]{inputenc}
\usepackage{graphicx}
\usepackage{amsmath}
\usepackage{amsfonts}
\usepackage{amssymb}
\usepackage{bbold}
\usepackage[usenames,dvipsnames,svgnames,table]{xcolor}
\usepackage[USenglish]{babel}
\usepackage{hyperref}
\usepackage{grffile}

\DeclareMathOperator{\Tr}{Tr\,}

\begin{document}

\title{Inflated nodes and surface states in superconducting
half-Heusler compounds}

\author{C. Timm}
\email{carsten.timm@tu-dresden.de}
\affiliation{Institute of Theoretical Physics,
Technische Universit\"at Dresden, 01062 Dresden, Germany}

\author{A. P. Schnyder}
\affiliation{Max-Planck-Institut f\"ur Festk\"orperforschung,
Heisenbergstrasse 1, 70569 Stuttgart, Germany}

\author{D. F. Agterberg}
\affiliation{Department of Physics, University of Wisconsin, Milwaukee,
Wisconsin 53201, USA}

\author{P. M. R. Brydon}
\email{philip.brydon@otago.ac.nz}
\affiliation{Department of Physics, University of Otago, P.O. Box 56,
Dunedin 9054, New Zealand}

\date{July 17, 2017}

\begin{abstract}
Two topics of high current interest in the field of unconventional
superconductivity are noncentrosymmetric superconductors and
multiband superconductivity. Half-Heusler superconductors such as
YPtBi exemplify both. In this paper, we study bulk and
surface states in nodal superconducting phases of the
half-Heusler compounds, belonging to the $A_1$ ($s+p$-like) and $T_2$
($k_zk_x+ik_zk_y$-like) irreducible
representations of the point group. These two phases preserve and break
time-reversal symmetry, respectively. For the $A_1$ case, we find that
flat surface bands persist in the multiband system. In addition, the
system has dispersive surface bands with zero-energy crossings forming
Fermi arcs, which are protected by mirror symmetries. For the $T_2$
case, there is an interesting coexistence of point and line nodes,
known from the single-band case, with Bogoliubov Fermi surfaces
(two-dimensional nodes). There are no flat-band surface states, as
expected, but dispersive surface bands with Fermi arcs exist. If these
arcs do not lie in high-symmetry planes, they are split by the
antisymmetric spin-orbit coupling so that their number is doubled
compared to the inversion-symmetric case. 
\end{abstract}

\maketitle

\section{Introduction}
\label{sec.introduction}

Topologically nontrivial superconducting states are currently
receiving a lot of attention, in part motivated by the vision of
topologically protected quantum computation \cite{NSS08}. Several
different approaches concern heterostructures in which topological
superconductivity is induced at the interface
\cite{Ali12,Bee13,StT13,NDL14}. The present paper belongs to another
research direction, which focuses on superconductors that are
intrinsically topologically nontrivial. For fully gapped
superconductors in any number of dimensions, the possible topological
states can be classified \cite{SRF08,Kit09} on the basis of the ten
Altland-Zirnbauer symmetry classes \cite{Zir96}. However, many
unconventional superconductors do not have a full gap but are nodal,
i.e., they feature quasi-particle states at the Fermi energy. It has
been realized that such superconductors can also be topologically
nontrivial and topological invariants associated with nodes of
dimension zero (points), one (lines), and two (surfaces) have been
constructed~\cite{RyH02,Vol03,TMY10,STY11,Vol11,BST11,ScR11,CTS16,ABT17,BzS17}.

An important class of unconventional superconductors are the
noncentrosymmetric materials \cite{BaS12,SSYA17}, in which the
absence of inversion symmetry allows  for the appearance
of antisymmetric spin-orbit coupling (ASOC).
In single-band systems, due to the ASOC, the spin of a Cooper pair is not a good
quantum number, leading to the mixing of (time-reversal-invariant)
singlet and triplet pairing. If the triplet component is sufficiently
large, the superconducting gap develops line nodes that are
topologically protected by an integer ($\mathbb{Z}$) winding number
and are accompanied by flat zero-energy surface bands
\cite{TMY10,STY11,BST11,ScR11,SBT12,ScB15}. Lattice
symmetries can induce additional topological invariants protecting
points and lines of zeros of the surface-state dispersion
\cite{SBT12,ScB15}. 
 
A particularly promising candidate in this class is the half-Heusler
compound YPtBi. Measurements of the upper critical field
versus temperature indicate a sizable triplet component
\cite{BNH12}. This is consistent with the observed linear temperature
dependence of the London penetration depth \cite{KWN16}, which is
attributed to line nodes. Moreover, tunneling spectra between a normal
conductor and superconducting YPtBi show a pronounced, though
broadened, zero-bias peak \cite{KWN16}, which agrees with expectations
for an extended flat zero-energy surface band
\cite{TMY10,STY11,BST11,ScR11,SBT12,ScB15}. This motivates the present
study of the surface dispersion of two superconducting states that
conventionally are expected to have line nodes. The possible pairing
states can be classified according to irreducible representations (irreps)
of the crystallographic point group $T_d$. The two states we consider are
a time-reversal-symmetric $A_1$, $s+p$-like pairing state with line nodes
\cite{KWN16,BWW16} and a $T_2$, $k_zk_x+ik_zk_y$-like pairing state that
breaks time-reversal symmetry and, in the limit of infinitesimal gaps, has
both point and line nodes \cite{BWW16}.

The $T_2$ state with broken time-reversal symmetry is also
interesting from the perspective of nodal excitations. In particular, it
has been shown that when inversion symmetry is present, this state
exhibits topologically protected nodal Bogoliubov Fermi surfaces
\cite{CTS16,ABT17,BzS17}, i.e., two-dimensional Fermi surfaces of
neutral Bo\-go\-liu\-bov quasiparticles. The inversion symmetry is required for
the topological protection of these Fermi surfaces
\cite{CTS16,ABT17,BzS17}, and their fate is unknown once this
protection is removed, as is the case in YPtBi. Here we show that with
noncentrosymmetric $T_d$ symmetry, this state exhibits a fascinating
coexistence of point nodes, line nodes, and Bogoliubov Fermi surfaces. 

Another very interesting aspect of the half-Heusler superconductors is
their topologically nontrivial normal state.
Due to the absence of inversion symmetry in the tetrahedral
point group $T_d$ and the resulting ASOC,
the degeneracy of energy bands is lifted, except at high-symmetry points in the
Brillouin zone. The most relevant band here is the four-component
$\Gamma_8$ band. Among the large family of half-Heusler compounds,
some show band inversion of the $\Gamma_8$ and the two-component
$\Gamma_6$ bands
\cite{CQK10,LWX10}. The Fermi energy for the undoped compounds then
lies at the $\Gamma_8$ point, assuming there is no accidental band
overlap away from the $\Gamma$ point. The compounds with inverted
bands can thus be viewed as semimetals with a quadratic band touching
point at the Fermi energy or as zero-gap semiconductors. Due to the
inverted bands, topologically protected surface states are expected
\cite{LYW16}. Band-structure calculations within density-functional
theory (DFT) predict the band inversion to be particularly large in
YPtBi and LuPtBi \cite{CQK10,LWX10}. Bands of dispersive surface
states have indeed been observed for YPtBi and LuPtBi using
angle-resolved photoemission spectroscopy (ARPES)
\cite{LYW16}. Superconductivity occurs both in YPtBi \cite{BSK11} and
in LuPtBi \cite{TFJ13}, with transition temperatures of
$T_c=0.77\,\mathrm{K}$ and $T_c=1.0\,\mathrm{K}$, respectively. 
The irreducible multiband character of these compounds due to the
dominant $\Gamma_8$ band should have interesting consequences for
superconductivity. Some of the authors have recently shown that
multiband systems allow for novel pairing states that are impossible
for a single band \cite{BWW16,ABT17,KWN16}.  

Nontrivial surface states are characteristic for topological materials
and provide the most important route to the experimental verification
of the topological state, e.g., through tunneling spectroscopy and
ARPES. Since superconducting YPtBi and LuPtBi are promising
candidates, it is worthwhile to study their surface states in some
detail. Not only should nontrivial superconductivity have signatures
in the surface dispersion but also the question arises as to what
happens to the surface bands of the normal state. 
In this paper, we analyze the bulk and surface dispersion of
half-Heusler superconductors. For numerical calculations, we take
parameters appropriate for YPtBi.
As mentioned above, we consider two representative pairing states:
a time-reversal-symmetric $A_1$
pairing state with line nodes \cite{KWN16,BWW16} and a $T_2$
pairing state that breaks time-reversal symmetry \cite{BWW16}.
Since the bands, including the surface bands, are nondegenerate it makes
sense to ask about the spin polarization of the Bloch
states. Specifically, we obtain the spin polarization of the surface
states for the $A_1$ state. 

The rest of this paper is organized as follows. In
Sec.\ \ref{sec.model}, we introduce the normal-state tight-binding
Hamiltonian for the half-Heusler compounds and then the Bogoliubov-de
Gennes Hamiltonian for their superconducting states. We give the
Hamiltonians both for the extended system and for slabs with (111) and
(100) surfaces. In Sec.\ \ref{sec.results}, we present and discuss our
results for the bulk and the surface, for the $A_1$ and $T_2$ pairing
states. Finally, we give a summary and draw conclusions in
Sec.~\ref{sec.concl}

\section{Model}
\label{sec.model}

We start by setting up an effective tight-binding model on the fcc
lattice, which is the Bravais lattice of the half-Heusler
structure. The edge length of the conventional, nonprimitive fcc unit
cell is set to $2$, the nearest-neighbor separation on the fcc lattice
is then $\sqrt{2}$. The four-component electron field of the
$\Gamma_8$ band is described in terms of an effective angular momentum
of $j=3/2$ \cite{FBG15,BWW16,YLW16,ABT17,KMH17,BoH17,SRV17}.
This angular momentum is due to
the coupling of the electron spins with $l=1$ \textit{p}-orbitals of
the main-group \textit{Z} ion in the half-Heusler materials with sum
formula \textit{XYZ}, in the present case Bi. The same model applies
to materials with zinc-blende structure. As noted in
Ref.\ \cite{ABT17}, it can also be used to formally describe
the four-component electronic fields generated by  two orbitals
and spin $1/2$. The mapping between the $j=3/2$ representation and the
orbital-spin representation is given in Ref.~\cite{ABT17}. 

Restricting ourselves to only nearest-neighbor hopping on the fcc
lattice, the normal-state Hamiltonian $H_N$ is given as a bilinear
form of the four-component spinor operator $c_i =
(c_{i,3/2},c_{i,1/2},c_{i,-1/2},c_{i,-3/2})^T$
($T$ denotes transposition) and its Hermitian conjugate
$c_i^\dagger$. The coefficients are expressed in terms of the
standard $4\times 4$ angular-momentum $j=3/2$ matrices $J_x$, $J_y$,
and $J_z$. The specific form in real space is
\begin{align}
H_N &= \sum_{ij} c_i^\dagger\, h_{ij}\, c_j
  = - t_1 \sum_{\langle ij\rangle}\, \big(c_i^\dagger c_j + \mathrm{H.c.}\big)
  \nonumber \\
& {}- t_2 \sum_{\langle ij\rangle} \left( c_i^\dagger J_{\eta_{ij}}^2 c_j
    + \mathrm{H.c.} \right) \nonumber \\
& {}- t_3 \sum_{\langle ij\rangle} \bigg( c_i^\dagger
    \sum_{\nu\neq\nu'} r_{ij,\nu} J_\nu\, r_{ij,\nu'} J_{\nu'} \,
    c_j + \mathrm{H.c.} \bigg) \nonumber \\
& {}- t_4 \sum_{\langle ij\rangle}\, \big( i c_i^\dagger \,
  \mathbf{r}_{ij} \cdot \mathbf{K}\, c_j + \mathrm{H.c.} \big) \nonumber \\
& {}- t_5 \sum_{\langle ij\rangle} \bigg[ i c_i^\dagger \sum_\nu r_{ij,\nu}\,
    (r_{ij,\nu+1}^2 - r_{ij,\nu+2}^2)\, J_\nu\, c_j + \mathrm{H.c.} \bigg]
  \nonumber \\
& {}- \mu \sum_i c_i^\dagger c_i ,
\label{HN.real.3}
\end{align}
where $\mathbf{r}_{ij} \equiv \mathbf{R}_i - \mathbf{R}_j$ in terms of
the fcc lattice sites $\mathbf{R}_i$, 
$\eta_{ij}=x$, $y$, and $z$ for $\mathbf{r}_{ij}$ perpendicular to the $x$, $y$, and $z$
axes, i.e., lying in the $yz$, $zx$, and $zx$ planes, respectively,
$\mathbf{K}$ is the
vector of matrices $K_\nu \equiv J_{\nu+1} J_\nu J_{\nu+1} - J_{\nu+2}
J_\nu J_{\nu+2}$, where the notation ``$\nu+1$'' and ``$\nu+2$'' pertains to
the cyclic group $\{x,y,z\}$, and $\sum_{\langle ij\rangle}$ denotes a
sum over nearest-neighbor bonds, counting each bond once. This
Hamiltonian is compatible with the space group $F\bar{4}3m$ of
half-Heusler compounds. The $t_4$ and $t_5$ terms represent the ASOC.
Our numerical results are obtained for $t_5=0$ and we drop the $t_5$
term from now on. However, all statements on symmetries and topological
protection remain valid in the presence of this term.

The superconducting state is described by the second-quantized Hamiltonian
\begin{equation}
H = \frac{1}{2} \sum_{ij} \Psi_i^\dagger\,
  \mathcal{H}_{ij}\, \Psi_j ,
\label{H.super.1}
\end{equation}
in terms of the Nambu spinor
\begin{equation} 
\Psi_i =
\left(\begin{array}{cc}
  c_i \\ c_i^{\dagger T}
  \end{array}\right)
\end{equation}
and the Bogoliubov-de Gennes Hamiltonian
\begin{equation}
{\mathcal H}_{ij} = \left(\begin{array}{cc}
  h_{ij} & \Delta_{ij} \\[1ex]
  \Delta_{ji}^\dagger & -h_{ji}^T
  \end{array}\right) ,
\label{H.super.1a}
\end{equation}
where $h_{ij}$ is defined in Eq.\ (\ref{HN.real.3}).
In this paper, we are not concerned with the origin of the
superconducting pairing. This has recently been considered
in Refs.~\cite{Mei16,BoH17,SRV17}.

The effective angular momentum $j=3/2$ allows for Cooper pairs with
total angular momenta $J=0$ (singlet), $J=1$ (triplet), $J=2$
(quintet), $J=3$ (septet) \cite{FBG15,BWW16,YLW16,BoH17,SRV17}.
By subducing the irreps of $\mathrm{O}(3)$ of order $J$ to the point
group $T_d$ and reducing them into irreps of $T_d$, one finds the
appropriate irreps for any $J$. In this way, one finds that, for a
purely local (\textit{s}-wave) pairing potential, there must be one 
singlet pairing state transforming according to the trivial irrep
$A_1$ and five quintet pairing states transforming according to the
two-dimensional irrep $E$ and to the three-dimensional irrep
$T_2$. The on-site pairing Hamiltonian can be written as
\begin{equation}
 H^{s}_{\mathrm{pair}}= \sum_{i,r} \left( \Delta^{0*}_r\, c_i^T \Gamma_r^\dagger c_i
  + \Delta^0_r\, c_i^\dagger \Gamma_r c_i^{\dagger T}\right),
\end{equation}
where the index $r$ enumerates the possible local pairing terms,
specifying the irrep and also an index counting
components for the higher-dimensional irreps.
The matrices $\Gamma_r=D_r U_T$ can be written as products of
irreducible tensor operators $D_r$ of the appropriate irreps
\cite{Rac42,Ste52,DaL72} and the unitary part
 \begin{equation}
U_T = \left(\begin{array}{cccc}
    0 & 0 & 0 & 1 \\
    0 & 0 & -1 & 0 \\
    0 & 1 & 0 & 0 \\
    -1 & 0 & 0 & 0
  \end{array}\right)
 \end{equation}
of the antiunitary time-reversal operator $\mathcal{T} = U_T\mathcal{K}$
($\mathcal{K}$ is complex conjugation). We hence find the matrices
\begin{align}
\Gamma_{A_1} &= U_T , \\
\Gamma_{E,1} &= \frac{1}{3}\, (2J_z^2 - J_x^2 - J_y^2)\, U_T , \label{Gamma.E.1} \\
\Gamma_{E,2} &= \frac{1}{\sqrt{3}}\, (J_x^2 - J_y^2)\, U_T , \\
\Gamma_{T_2,1} &= \frac{1}{\sqrt{3}}\, (J_y J_z + J_z J_y)\, U_T , \\
\Gamma_{T_2,2} &= \frac{1}{\sqrt{3}}\, (J_z J_x + J_x J_z)\, U_T , \\
\Gamma_{T_2,3} &= \frac{1}{\sqrt{3}}\, (J_x J_y + J_y J_x)\, U_T . \label{Gamma.T2.3}
\end{align}
$\Gamma_{A_1}$ belongs to the singlet ($J=0$) irrep $A_1$, whereas the
other five belong to the quintet ($J=2$) irreps $E$ and $T_2$. All six
matrices are invariant under time reversal.
This local pairing enters $\Delta_{ij}$ in Eq.\ (\ref{H.super.1a}) as
\begin{equation}
\Delta^s_{ij} = 2 \delta_{ij} \sum_r \Delta^0_r\, \Gamma_r .
\end{equation}
A local pairing term that transforms according to a certain irrep will
generically be accompanied by nonlocal pairing transforming in the
same manner. The best studied example is singlet-triplet mixing in
single-band noncentrosymmetric superconductors
\cite{Sat06,Ber10,BST11,ScR11,SBT12,ScB15}. Frigeri \textit{et
al.}\ \cite{FAK04} have shown for the single-band case that pair
breaking due to the ASOC is avoided for triplet pairing with the same
momentum (and hence spatial) dependence as the ASOC. This is thus
expected to be the most favorable triplet pairing state. 

The natural generalization to the $j=3/2$ case is the pairing matrix
$\Delta^p_{ij} \propto h_{ij}^\mathrm{ASOC}\, U_T$, where
$h_{ij}^\mathrm{ASOC}$ is the ASOC part of $h_{ij}$. The superscript
\textit{p} stands for \textit{p}-wave. We thus write
\begin{equation}
\Delta^p_{ij} = -i\, \Delta^0_p\, \mathbf{r}_{ij} \cdot \mathbf{K}\, U_T ,
\label{Delta.pij.3}
\end{equation}
where $i$ and $j$ are nearest-neighbor sites. This pairing term satisfies time-reversal symmetry. By construction, it is also invariant under the lattice symmetries of the normal state. For this reason, it will generically coexist with the local $A_1$ singlet pairing, which also transforms trivially under the lattice symmetries \cite{BWW16}.

The real-space formulation can be used for both extended, bulk system and for slabs of various orientations. We now discuss these two cases in turn.

\subsection{Extended system}

For the extended system, we Fourier transform the Hamiltonian in all
three directions. The normal-state tight-binding Hamiltonian is then
\begin{equation}
  H_N= \sum_\mathbf{k} c_\mathbf{k}^\dagger\, h(\mathbf{k})\, c_\mathbf{k} ,
\end{equation}
where
\begin{align}
h(\mathbf{k})
&=  -4t_1 \sum_{\nu} \cos k_\nu \cos k_{\nu+1} \nonumber \\
& {}- 4t_2 \sum_{\nu} \cos k_{\nu} \cos k_{\nu+1}\, J_{\nu+2}^2 \nonumber \\
& {}+ 4t_3 \sum_{\nu}  \sin k_{\nu} \sin k_{\nu+1}\,
  (J_\nu J_{\nu+1} + J_{\nu+1}J_\nu)  \nonumber \\
& {}- 4t_4 \sum_{\nu} \sin k_\nu\,(\cos k_{\nu+1}+\cos k_{\nu+2})\, K_\nu
 - \mu
\label{normal.HTBk.3}
\end{align}
and  $c_\mathbf{k}=(c_{\mathbf{k},3/2},c_{\mathbf{k},1/2},
c_{\mathbf{k},-1/2},c_{\mathbf{k},-3/2})^T$ is the four-com\-po\-nent spinor
operator. The momentum sum is over the fcc Brillouin zone. The energy of the
$\Gamma_8$ band-tou\-ching point is $E^N_0 = -12t_1 - 15t_2$. The
expansion of the coefficients for small $\mathbf{k}$ gives the
$\mathbf{k}\cdot\mathbf{p}$ Hamiltonian 
\begin{align}
h_{\mathbf{k}\cdot\mathbf{p}}(\mathbf{k})
  &= \alpha k^2 + \beta \sum_\nu k_\nu^2 J_\nu^2
  + \gamma \sum_{\nu\neq\nu'} k_\nu k_{\nu'} J_\nu J_{\nu'} \nonumber \\
& {}+ \delta \sum_\nu k_\nu K_\nu - \tilde\mu ,
\label{H.kp}
\end{align}
where $t_1=\alpha/4+15\beta/16$, $t_2=-\beta/2$, $t_3=\gamma/4$,
$t_4=-\delta/8$, and $\mu = \tilde\mu - 3\alpha - 15\beta/4$.
We set $\hbar=1$ throughout this paper. Identity matrices
are suppressed in the first and last terms. This is the
$\mathbf{k}\cdot\mathbf{p}$ Hamiltonian used in Ref.~\cite{BWW16}.

Brydon \textit{et al.}\ \cite{BWW16} have performed band structure calculations for YPtBi and LuPtBi within DFT, using various approximations for the exchange-correlation functional. The results show significant quantitative differences but agree on the band topology, in particular on the band inversion. The modified Becke-Johnson local-density approximation \cite{TrB09}, which was developed to improve the calculated band gaps, predicts stoichiometric YPtBi to be a semimetal with its Fermi energy at the quadratic band-touching point \cite{BWW16}.

To be specific, for YPtBi we take the same parameters as in \cite{BWW16}, which correspond to $t_1=-0.918\,\mathrm{eV}$, $t_2=0.760\,\mathrm{eV}$, $t_3=-0.253\,\mathrm{eV}$, and $t_4=-3.98\,\mathrm{meV}$. These parameters give reasonable quantitative agreement with the results of the modified Becke-Johnson local-density approximation at small momenta $\mathbf{k}$ and avoids spurious large Fermi surfaces away from the $\Gamma$ point. Inclusion of the cubic ASOC ($t_5$) does not significantly improve the agreement. The chemical potential is taken to be $\mu = E^N_0 - 0.02\,\mathrm{eV}$, corresponding to weak hole doping, as seen in experiments~\cite{BSK11,BNH12}. Note that the description of LuPtBi, which has a much more complex normal-state Fermi surface, would require longer-range hoppings in the Hamiltonian.

The four eigenenergies in momentum space are generically
nondegenerate. They are strongly split into two pairs by the symmetric
$t_2$ and $t_3$ terms. In YPtBi, one pair of bands curves up and the
other down. The pairs are more weakly split by the antisymmetric $t_4$
term. All four bands are degenerate at $\mathbf{k}=0$. This
band-touching point is protected by a combination of time-reversal and
lattice symmetries: splitting it would require a (mass) term in the
Hamiltonian that is independent of $\mathbf{k}$ and not proportional
to the identity matrix. One can easily convince oneself that there are
only six linearly independent $4\times 4$ matrices that are Hermitian
and satisfy time-reversal symmetry. A possible choice for the six
matrices is $J_x^2$, $J_y^2$, $J_z^2$, $J_xJ_y+J_yJ_x$,
$J_yJ_z+J_zJ_y$, and $J_zJ_x+J_xJ_z$. One can then check that there is
no combination except for the identity matrix that also satisfies all
symmetries of the point group $T_d$. 

For weak electron or hole doping, the model develops two small, nested Fermi pockets. The pockets have first-order (conical) touching points on the $k_x$, $k_y$, and $k_z$ axes, which are protected by the twofold degeneracy of bands on these axes following from the splitting of the four-dimensional irrep $\Gamma_8$ of $T_d$ into two-dimensional irreps $\Delta_5+\Delta_5$ of $C_{2v}$.

The superconducting system is described by the Hamiltonian
\begin{equation}
H = \frac{1}{2} \sum_\mathbf{k} \Psi_\mathbf{k}^\dagger\,
  {\mathcal H}(\mathbf{k}) \, \Psi_\mathbf{k} ,
\label{H.super.3}
\end{equation}
in terms of the Nambu spinor $\Psi_\mathbf{k} = (c_{\mathbf{k}}^T,
c^\dagger_{-\mathbf{k}})^T$, 
and the Bogoliubov-de Gennes Hamiltonian
\begin{equation}
\mathcal{H}(\mathbf{k}) = \left(\begin{array}{cc}
  h(\mathbf{k}) & \Delta(\mathbf{k}) \\[1ex]
  \Delta^\dagger(\mathbf{k}) & -h^T(-\mathbf{k})
  \end{array}\right) .
\label{H.super.3a}
\end{equation}
Here, $h(\mathbf{k})$ is the normal-state Hamiltonian defined in Eq.\ (\ref{normal.HTBk.3}) and the pairing potential is $\Delta(\mathbf{k}) = \Delta^s(\mathbf{k}) + \Delta^p(\mathbf{k})$ with
\begin{align}
\Delta^s(\mathbf{k}) &= 2\sum_r \Delta^0_r \Gamma_r , \\
\Delta^p(\mathbf{k}) &= -4\Delta^0_p\, \sum_{\nu} \sin k_{\nu}\,
(\cos k_{\nu+1}+\cos k_{\nu+2})\, K_{\nu} U_T .
\label{Delta.3a}
\end{align}
Note that the \textit{p}-wave gap $\Delta^p({\bf k})$ is only present in the $A_1$ state.

\subsection{Slabs of finite thickness}

In the following section, we will show results of the numerical diagonalization of the Bogoliubov-de Gennes tight-binding model on slabs with (111) and with (100) surface orientations. The thickness $W$ of the slabs has to be chosen large enough to suppress the hybridization of states localized at opposite surfaces. Since the slabs have translation symmetry in the directions parallel to the surfaces, we block diagonalize the Hamiltonian in Eq.\ (\ref{HN.real.3}) by performing a Fourier transformation in these two directions. The corresponding wave vector parallel to the surfaces is denoted by $\mathbf{k}_\|=(k_1,k_2)$.

\begin{figure}[htb]
\begin{center}
\raisebox{1ex}{(a)}\includegraphics[width=0.65\columnwidth]{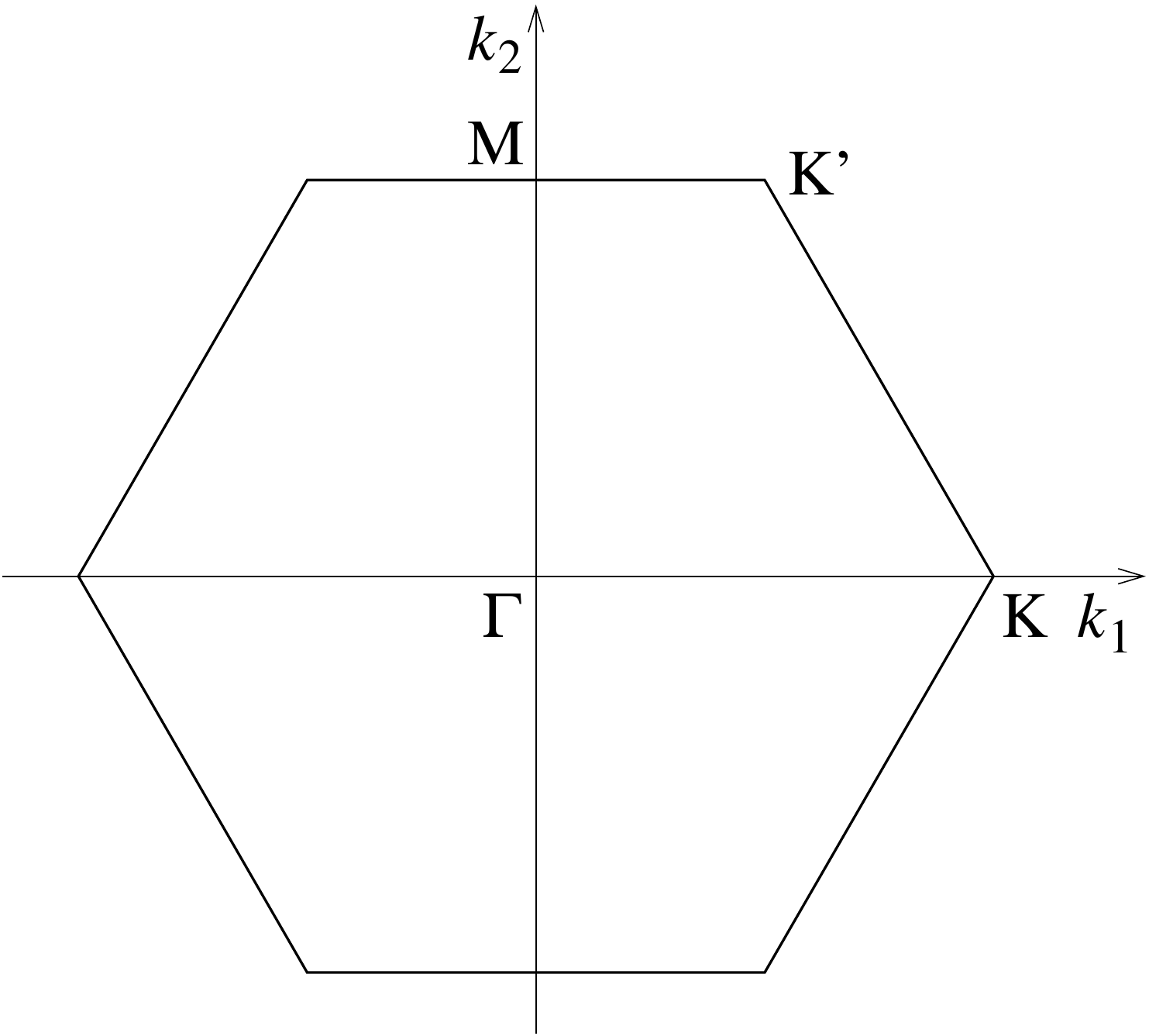}\\[2ex]%
\raisebox{1ex}{(b)}\includegraphics[width=0.65\columnwidth]{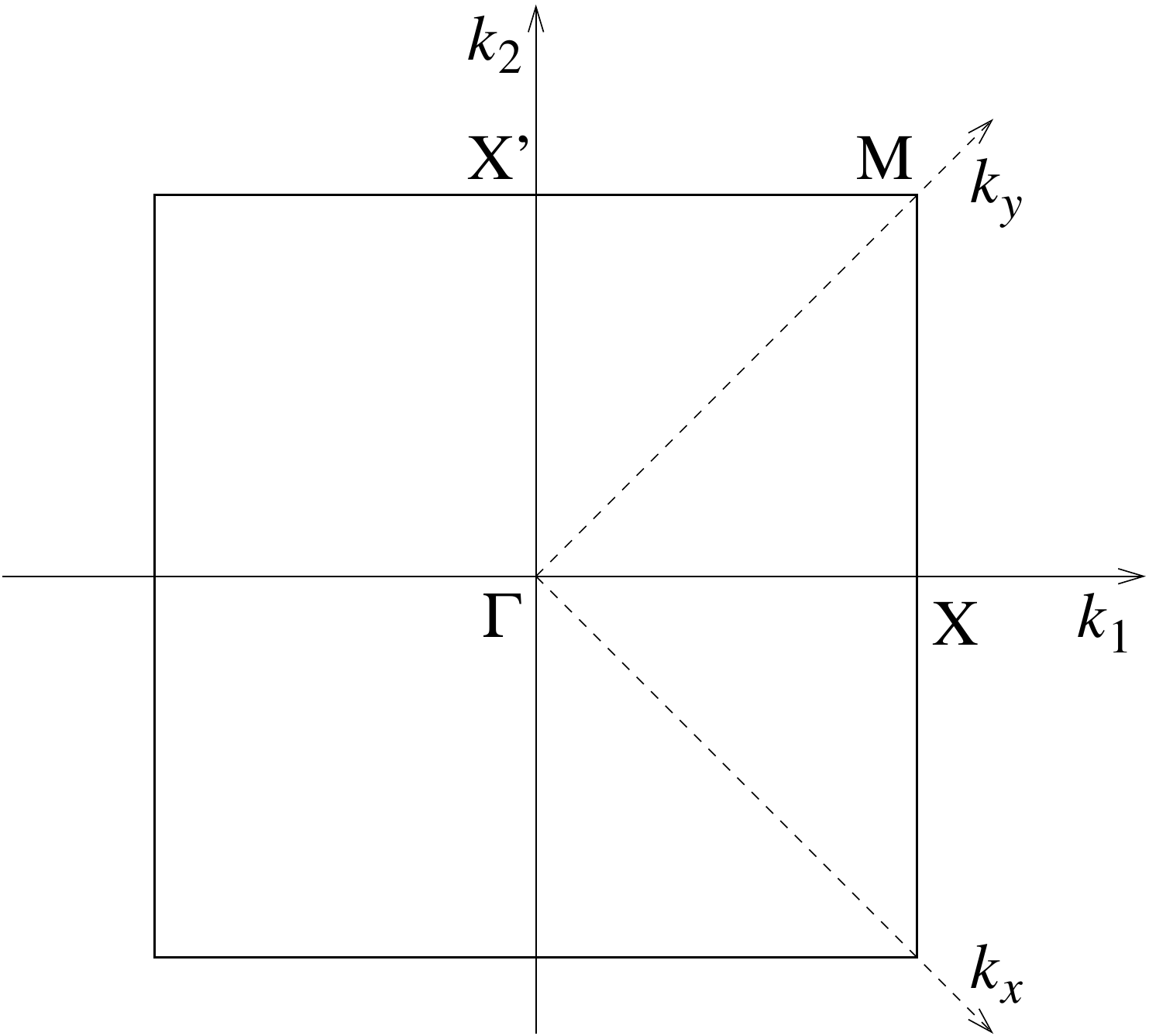}
\end{center}
\caption{(a) Two-dimensional Brillouin zone of a slab with (a) (111) and (b) (100) surfaces.}
\label{fig.2DBZ}
\end{figure}

We start with a slab with (111) surfaces. The primitive bulk unit cell compatible with the symmetry of the (111) slab is hexagonal and contains three fcc sites. The thickness $W$ is here defined as the number of triangular layers parallel to the surfaces, which means that the unit cell has a height of $3$ layers and the slab is $W/3$ hexagonal unit cells thick. The layers are enumerated by $l = 0, \ldots, W-1$. Momentum sums are taken over the two-dimensional Brillouin zone of the slab, with is a hexagon. The components $(k_1,k_2)$ are defined in Fig.\ \ref{fig.2DBZ}(a). Special points have the two-dimensional coordinates $\mathrm{K} = (2\sqrt{2}\,\pi/3,0)$,
$\mathrm{K'} = (\sqrt{2}\,\pi/3,\sqrt{2/3}\,\pi)$, and
$\mathrm{M} = (0,\sqrt{2/3}\,\pi)$. In terms of the bulk coordinate system, the momentum reads
\begin{equation}
\mathbf{k}_\| = k_1\, \frac{1}{\sqrt{2}}\left(\begin{array}{@{}c@{}}
  1 \\ -1 \\ 0
  \end{array}\right) + k_2\, \frac{1}{\sqrt{6}}\left(\begin{array}{@{}c@{}}
  1 \\ 1 \\ -2
  \end{array}\right) .
\label{normal.kpar.2}
\end{equation}
The second-quantized Hamiltonian for the slab is
\begin{equation}
H_\mathrm{slab} = \frac{1}{2} \sum_{\mathbf{k}_\|} \sum_{ll'}
  \Psi_{\mathbf{k}_\|l}^\dagger\, \mathcal{H}^{(111)}_{ll'}(\mathbf{k}_\|)\,
  \Psi_{\mathbf{k}_\|l'} ,
\label{H.slab.111.2}
\end{equation}
with the obvious definition of $\Psi_{\mathbf{k}_\|l}$ and the matrices
\begin{equation}
\mathcal{H}^{(111)}_{ll'}(\mathbf{k}_\|) = \left(\begin{array}{cc}
  h^{(111)}_{ll'}(\mathbf{k}_\|) & \Delta^{(111)}_{ll'}(\mathbf{k}_\|) \\[1ex]
  \Delta^{(111)}_{l'l}(\mathbf{k}_\|)^\dagger & -h^{(111)}_{l'l}(-\mathbf{k}_\|)^T
  \end{array}\right) ,
\label{H.slab.111.4}
\end{equation}
which can be expressed in terms of $4\times 4$ blocks $h_{ll'}^{(111)}(\mathbf{k}_\|)$
and $\Delta^{(111)}_{ll'}(\mathbf{k}_\|)$. The construction of these
blocks is a straightforward exercise and their explicit forms are omitted here.

We next consider a slab with (100) surfaces. The (nonprimitive) bulk unit cell compatible with the symmetry of this slab is centered tetragonal and contains two sites. The thickness $W$ is defined as the number of square layers parallel to the surfaces, which means that the slab is $W/2$ tetragonal unit cells thick. The layers are again enumerated by $l=0,\ldots,W-1$.
The momentum vector parallel to the surface is
\begin{equation}
\mathbf{k}_\| = k_1\, \frac{1}{\sqrt{2}} \left(\begin{array}{@{}c@{}}
  0 \\ 1 \\ 1
  \end{array}\right) + k_2\, \frac{1}{\sqrt{2}}
  \left(\begin{array}{@{}c@{}}
  0 \\ -1 \\ 1
  \end{array}\right) .
\end{equation}
The two-dimensional Brillouin zone of the slab is a square, which is shown in Fig.\ \ref{fig.2DBZ}(b). Note that the $k_1$ and $k_2$ axes for the slab are rotated by $45^\circ$ with respect to the conventional cubic axes of the bulk. The two-dimensional coordinates $(k_1,k_2)$ of special points are $\mathrm{M}=(\pi/\sqrt{2},\pi/\sqrt{2})$, $\mathrm{X}=(\pi/\sqrt{2},0)$, and $\mathrm{X}'=(0,\pi/\sqrt{2})$. The Hamiltonian reads
\begin{equation}
H_\mathrm{slab} = \frac{1}{2} \sum_{\mathbf{k}_\|} \sum_{ll'}
  \Psi_{\mathbf{k}_\|l}^\dagger\, \mathcal{H}^{(100)}_{ll'}(\mathbf{k}_\|)\,
  \Psi_{\mathbf{k}_\|l'} ,
\label{H.slab.001.2}
\end{equation}
with
\begin{equation}
\mathcal{H}^{(100)}_{ll'}(\mathbf{k}_\|) = \left(\begin{array}{cc}
  h^{(100)}_{ll'}(\mathbf{k}_\|) & \Delta^{(100)}_{ll'}(\mathbf{k}_\|) \\[1ex]
  \Delta^{(100)}_{l'l}(\mathbf{k}_\|)^\dagger & -h^{(100)}_{l'l}(-\mathbf{k}_\|)^T
  \end{array}\right) .
\label{H.slab.001.4}
\end{equation}

To distinguish surface from bulk states, we use the total weight in all bands in the central third of the slab. This quantity shows a large contrast between the two types of states and only weak finite-size effects. For time-reversal-symmetric pairing, we take into account the two states with lowest energy by absolute value since the spectrum is symmetric about the Fermi energy.

\section{Results}
\label{sec.results}

The model system possesses surface states even in the normal phase. This was already found for related gapless semiconductors by D'yakonov and A. V. Khaetskii \cite{DyK81} and recently within DFT for $\mathrm{(Y,Lu)PtBi}$ \cite{LYW16}. It is worth pointing out that the DFT calculations \cite{LYW16} indicate that the dispersion of the surface bands depends on the termination of the surface, an aspect that is missed by our simple tight-binding model. In addition, DFT calculations as well as ARPES \cite{LYW16} show a topological Dirac cone of surface states located \emph{below} the hole-like $\Gamma_8$ band. These surface states connect the bulk $\Gamma_8$ bands to the $\Gamma_6$ bands below and are thus not captured by our $\Gamma_8$-only tight-binding model. That the Dirac cone indeed derives from these bands is shown by Chu \textit{et al.}\ \cite{CSL11}, who use a six-band continuum model containing the $\Gamma_8$ and $\Gamma_6$ bands.

The presence of surface states derived solely from the $\Gamma_8$ band
can be understood based on a deformation of the Hamiltonian into a
topologically nontrivial one. Such an argument can be used to explain
the states localized at zig-zag edges of graphene and also to predict
surface states at certain surfaces of iron pnictides \cite{LaT13}. The
procedure starts by fixing the wave-vector components $\mathbf{k}_\|$
parallel to the surface. This produces an effectively one-dimensional
model with coordinate $l$, for which the states in reciprocal space
are enumerated by a wave number $k_\perp$. We consider wave vectors
$\mathbf{k}_\|\neq 0$, for which this one-dimensional model is
gapped. The corresponding Hamiltonian is then deformed, without
closing the gap, into one that has additional symmetries and
is topologically nontrivial. We do not
give the details here since the manipulations are very similar to
Ref.\ \cite{LaT13}. We end up with two decoupled one-dimensional
Hamiltonians in Altland-Zirnbauer class BDI, which allows a
$\mathbb{Z}$ topological invariant \cite{SRF08,Kit09,CTS16}, which in
this case turns out to be $\pm 1$. Hence, the deformed model has two
zero-energy surface state per surface, i.e., four in total,
for each $\mathbf{k}_\|\neq
0$. Now reversing the deformation, the topological protection of these
surface states is lost but they evolve continuously as a function of the
deformation. Hence, for small deformations, surface states survive but are
no longer pinned to zero energy, nor do they remain degenerate. We
thus generically expect four surface states at
$\mathbf{k}_\|$-dependent energies, i.e., four dispersive surface
bands. These surface bands only vanish if the deformation is so large
that they become resonant with bulk states.  

\begin{figure}[tb]
\centerline{\includegraphics[width=0.9\columnwidth]{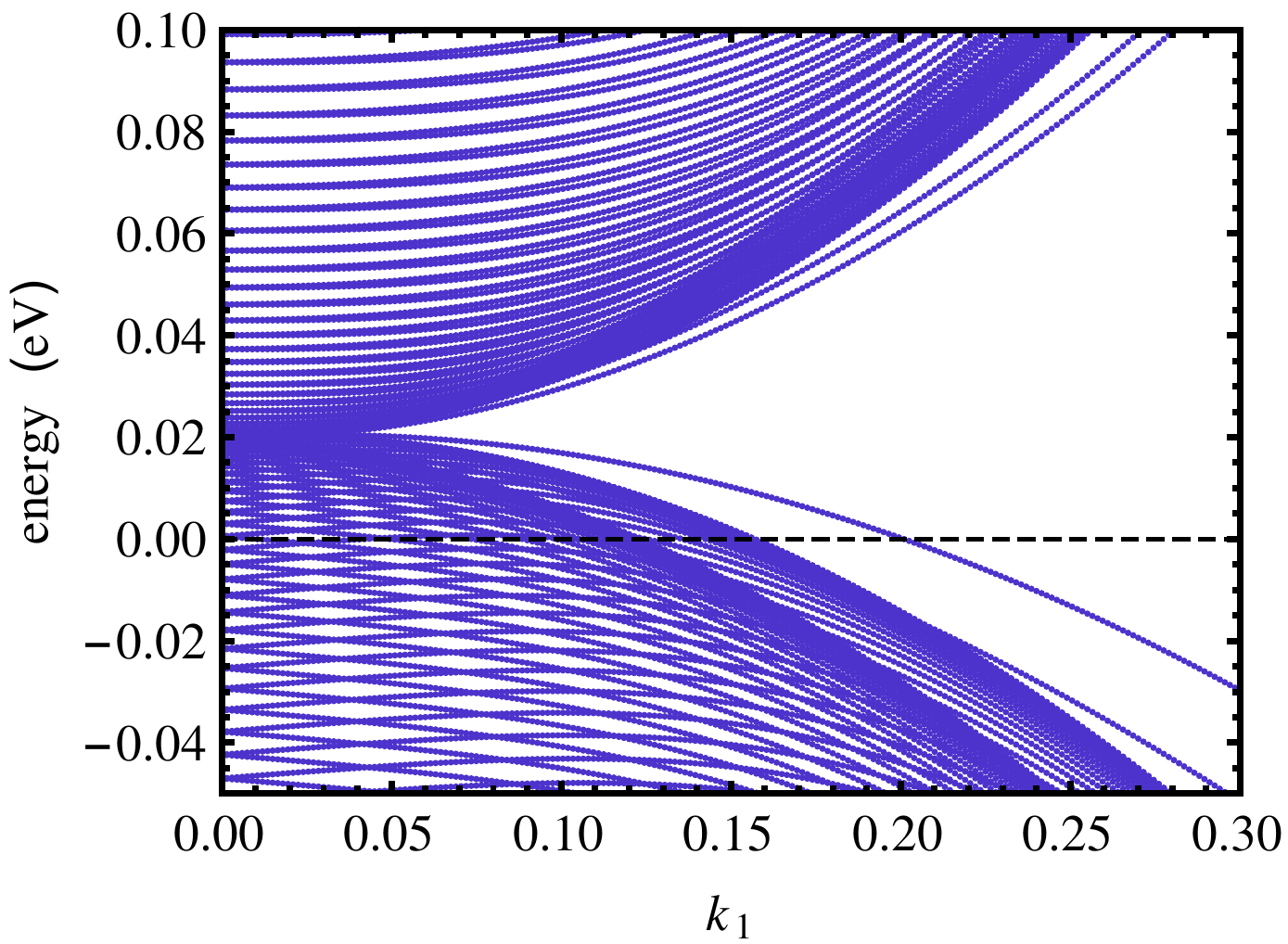}}
\caption{Band structure for the normal state of a slab with (100) surfaces along the $k_1$ axis. The thickness is $W=400$. The zero of energy is set to the Fermi energy (dashed horizontal line). The split-off surface bands are clearly visible.}
\label{fig.normal.001surface.k1}
\end{figure}

A cut through the band structure along the positive $k_1$ axis for a (100) slab in the normal state is plotted in Fig.\ \ref{fig.normal.001surface.k1}. The plot is restricted to energies close to the Fermi energy and to momenta in the region of the Fermi sea. The quasi-continuous regions correspond to bulk states that are weakly modified by the presence of surfaces. The surface bands are clearly visible in the gap between the bulk bands.

\begin{figure}[tb]
\begin{center}
\raisebox{0.5ex}{(a)}\includegraphics[width=0.95\columnwidth]{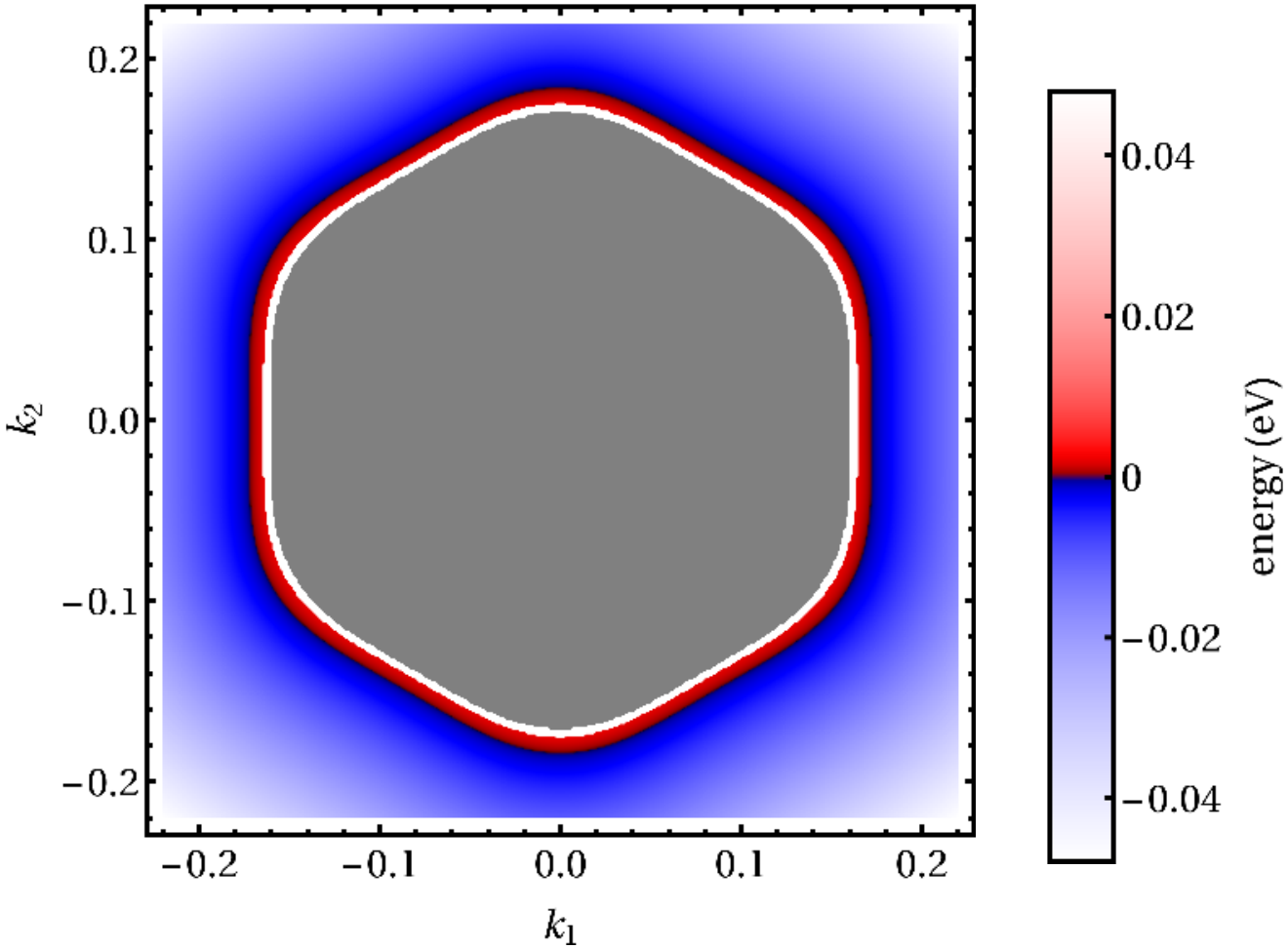}\\[2ex]%
\raisebox{0.5ex}{(b)}\includegraphics[width=0.95\columnwidth]{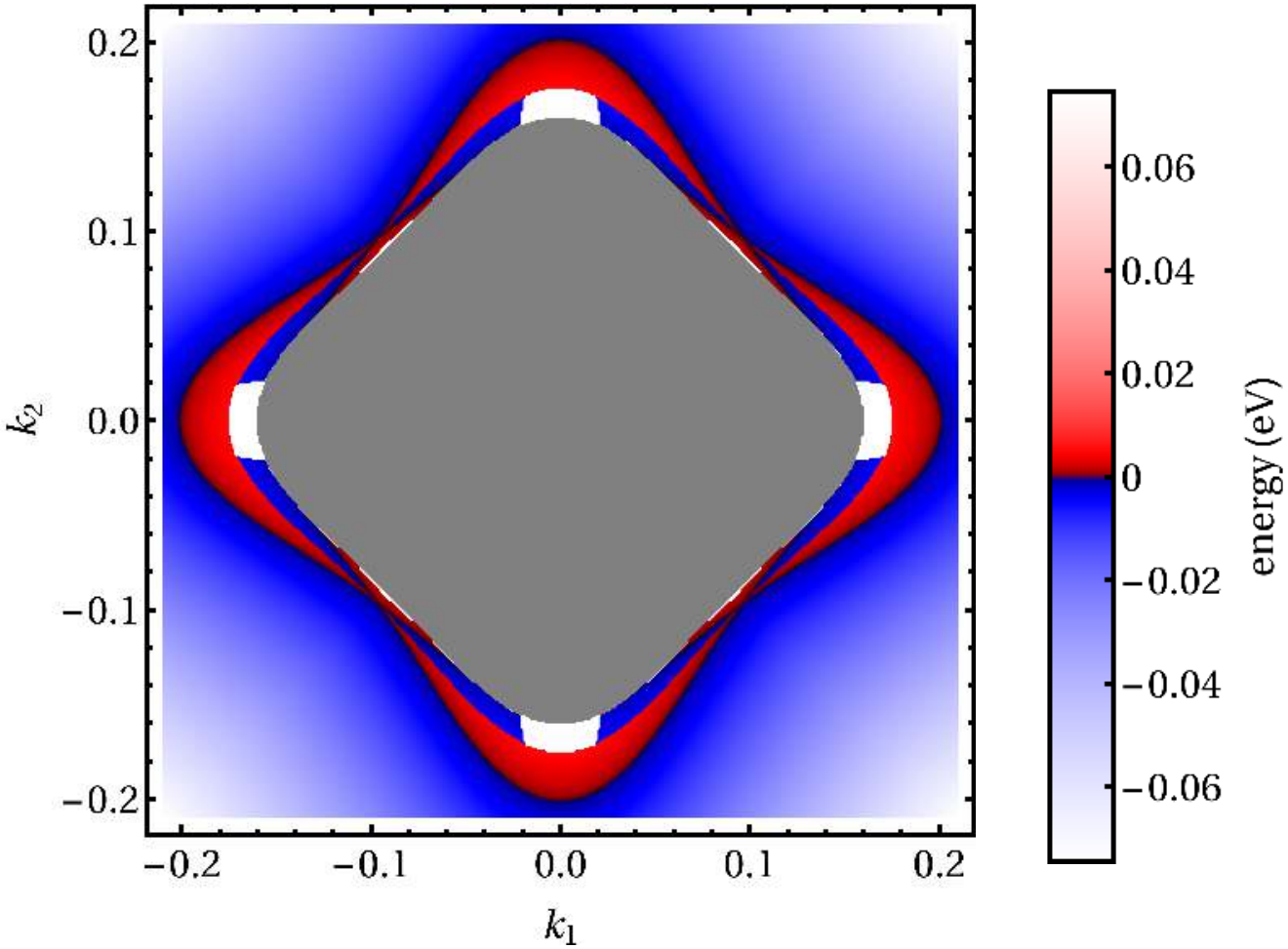}
\end{center}
\caption{Dispersion of surface states of YPtBi in the normal state for (a) the (111) surface and (b) the (100) surface, for a thickness of $W=2000$. The gray region in the center is the projection of the bulk Fermi sea, i.e., in this region the states at the Fermi energy are bulk like. Note that this does not preclude the presence of surface states away from the Fermi energy. In the white regions, there is no bulk Fermi sea but the states closest to the Fermi energy are still bulk like.}
\label{fig.normal.surface}
\end{figure}

Numerical results for the energy of surface states at (111) and (100)
surfaces in the normal phase are shown in
Fig.\ \ref{fig.normal.surface}. Here and in the following, we only
show the central region of the two-dimensional Brillouin zone, in the
vicinity of the normal-state Fermi sea. The plot only pertains to the
state closest to the Fermi energy for each $\mathbf{k}_\|$. These
states are extended through the bulk in the region of the projected
Fermi sea (gray in Fig.\ \ref{fig.normal.surface}) but also in small
regions outside of this projection (white). The color denotes the
energy of the surface state closest to the Fermi energy. These
correspond to the surface bands found by ARPES \cite{LYW16} and also
seen in Fig.\ \ref{fig.normal.001surface.k1}. In agreement with the
ARPES experiments and DFT calculations \cite{LYW16}, the surface bands
cross the Fermi energy from positive to negative values for increasing
momentum $\mathbf{k}_\|=(k_1,k_2)$. In Fig.\ \ref{fig.normal.surface},
this is indicated by a smooth change of color from red through black
to blue. 

The additional abrupt change of color seen for the (100) surface in
Fig.\ \ref{fig.normal.surface}(b) is an artifact of the presentation:
here, two surface bands have energies of the same absolute value but
opposite sign. 
The white regions in Fig.\ \ref{fig.normal.surface}(b) are not a
finite-size effect. Figure \ref{fig.normal.001surface.k1} shows
clearly what is happening here: the surface bands continue but the
states closest to the Fermi energy are now bulk like. At smaller
momentum, the continuum of bulk states reaches the Fermi energy,
corresponding to the gray region in Fig.\ \ref{fig.normal.surface},
but the surface bands still continue and approach the quadratic
band-touching point.

\subsection{$A_1$ pairing: flat-band surface states and mirror Fermi arcs}
\label{sec.A1}

\begin{figure}[tb]
\begin{center}
\raisebox{1ex}{(a)}\includegraphics[width=0.7\columnwidth]{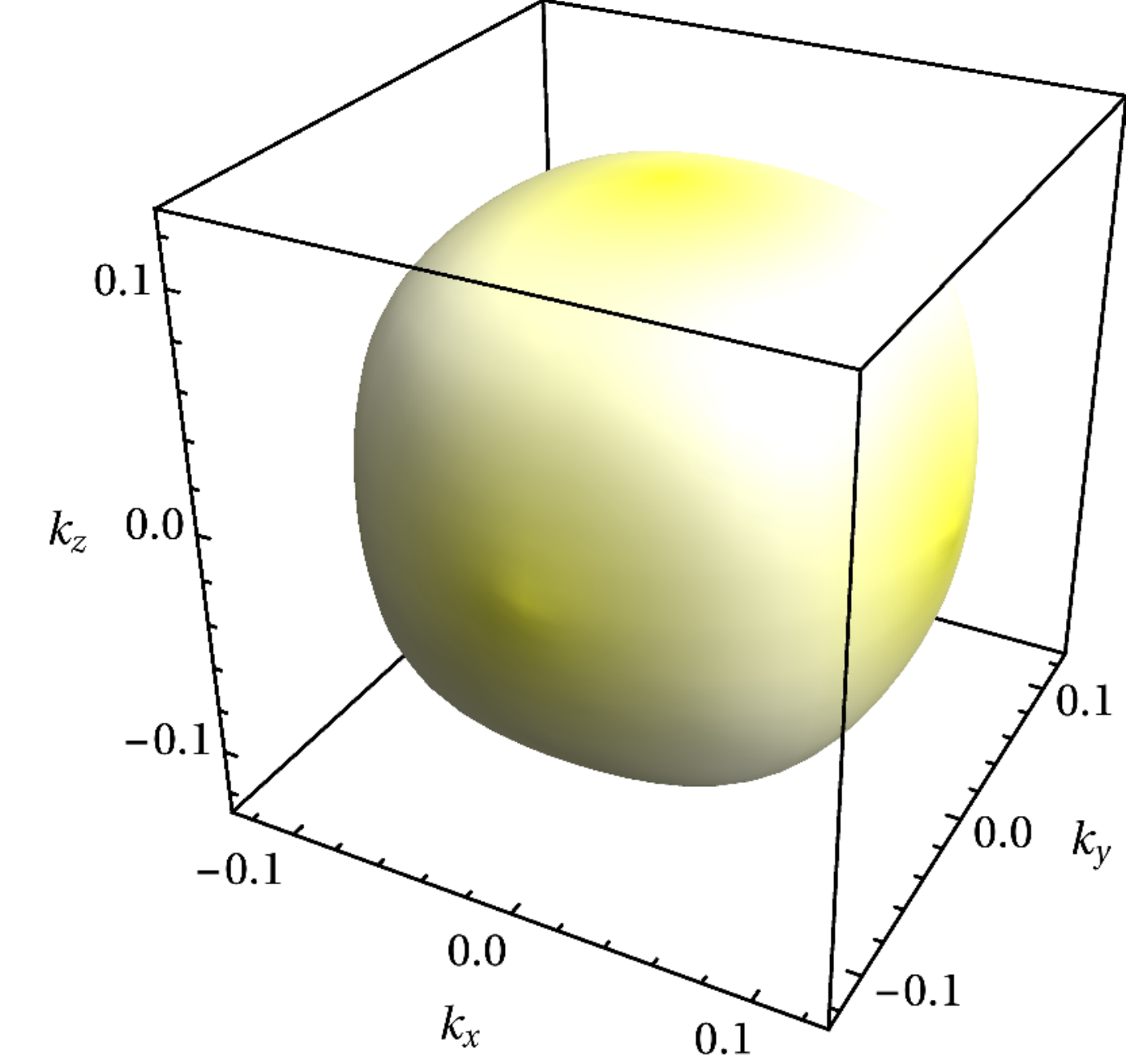}\\[2ex]%
\raisebox{1ex}{(b)}\includegraphics[width=0.7\columnwidth]{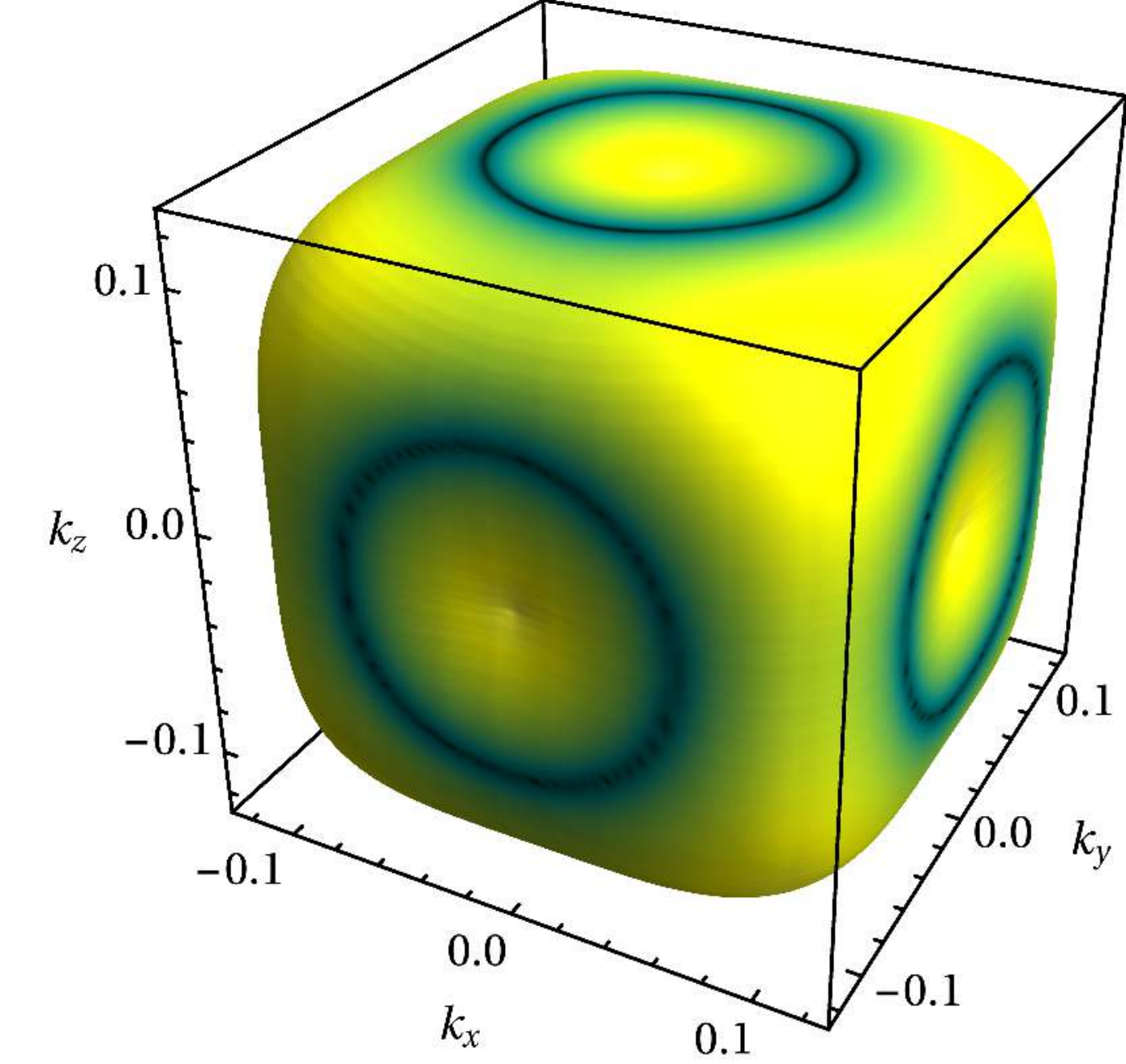}\\[2ex]%
\hspace{2em}\includegraphics[width=0.6\columnwidth]{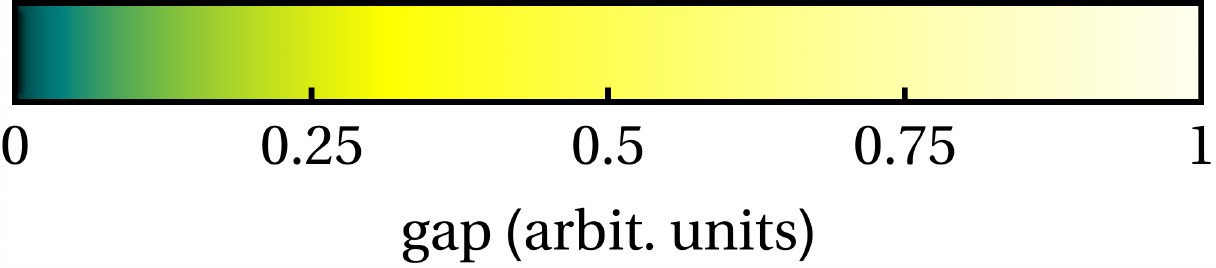}
\end{center}
\caption{Superconducting gap for the $A_1$ pairing state with \textit{p}-wave to \textit{s}-wave ratio $\Delta^0_p/\Delta^0_{A_1} = 7/3$ on (a) the smaller and (b) the larger normal-state Fermi surface. Infinitesimal pairing amplitudes have been used for reasons explained in the text.}
\label{fig.A1.gap}
\end{figure}

For the superconducting state, we first consider the $A_1$ gap matrix,
in real space,
\begin{align}
\Delta_{ij} &= 2 \delta_{ij}\, \Delta^0_{A_1} \Gamma_{A_1} 
  -i \delta_{\langle ij\rangle}\, \Delta^0_p\,
  \mathbf{r}_{ij}\cdot\mathbf{K} U_T ,
\end{align}
where $\delta_{\langle ij\rangle}$ is unity (zero) if $i$ and $j$ are
(not) nearest-neigh\-bor sides.
The amplitudes $\Delta^0_{A_1}$ and $\Delta^0_p$ are both taken to be
real. As noted above, both terms have the same, namely trivial,
transformation properties under all lattice symmetries and thus
generically coexist \cite{BWW16}. The state also preserves time-reversal
symmetry. The superconducting gap has line nodes when
$\Delta^0_p/\Delta^0_{A_1}$ is sufficiently large. We take
$\Delta^0_{A_1} = 3\,\mathrm{meV}$ and $\Delta^0_p = 7\,\mathrm{meV}$,
which leads to six closed line nodes on the larger Fermi surface,
surrounding the bulk cubic axes \cite{BWW16}. The smaller Fermi surface
is fully gapped. Inverting the sign of $\Delta^0_{A_1}$ or $\Delta^0_p$
moves the nodes to the smaller Fermi surface. Figure \ref{fig.A1.gap}
shows the absolute value of the gap on the normal-state Fermi surfaces
for infinitesimal pairing amplitudes with the same
\textit{p}-wave--to--\textit{s}-wave ratio $\Delta^0_p/\Delta^0_{A_1} = 7/3$.
The gap is obtained by treating the pairing in first-order perturbation
theory. Infinitesimal amplitudes are used here for illustration since for
larger amplitudes the energy minima and in particular the nodal rings move
away from the normal-state Fermi surfaces. However, the six nodal rings
persist for the larger amplitudes used in the following.

\begin{figure}[tb]
\begin{center}
\raisebox{0.5ex}{(a)}\includegraphics[width=0.95\columnwidth]{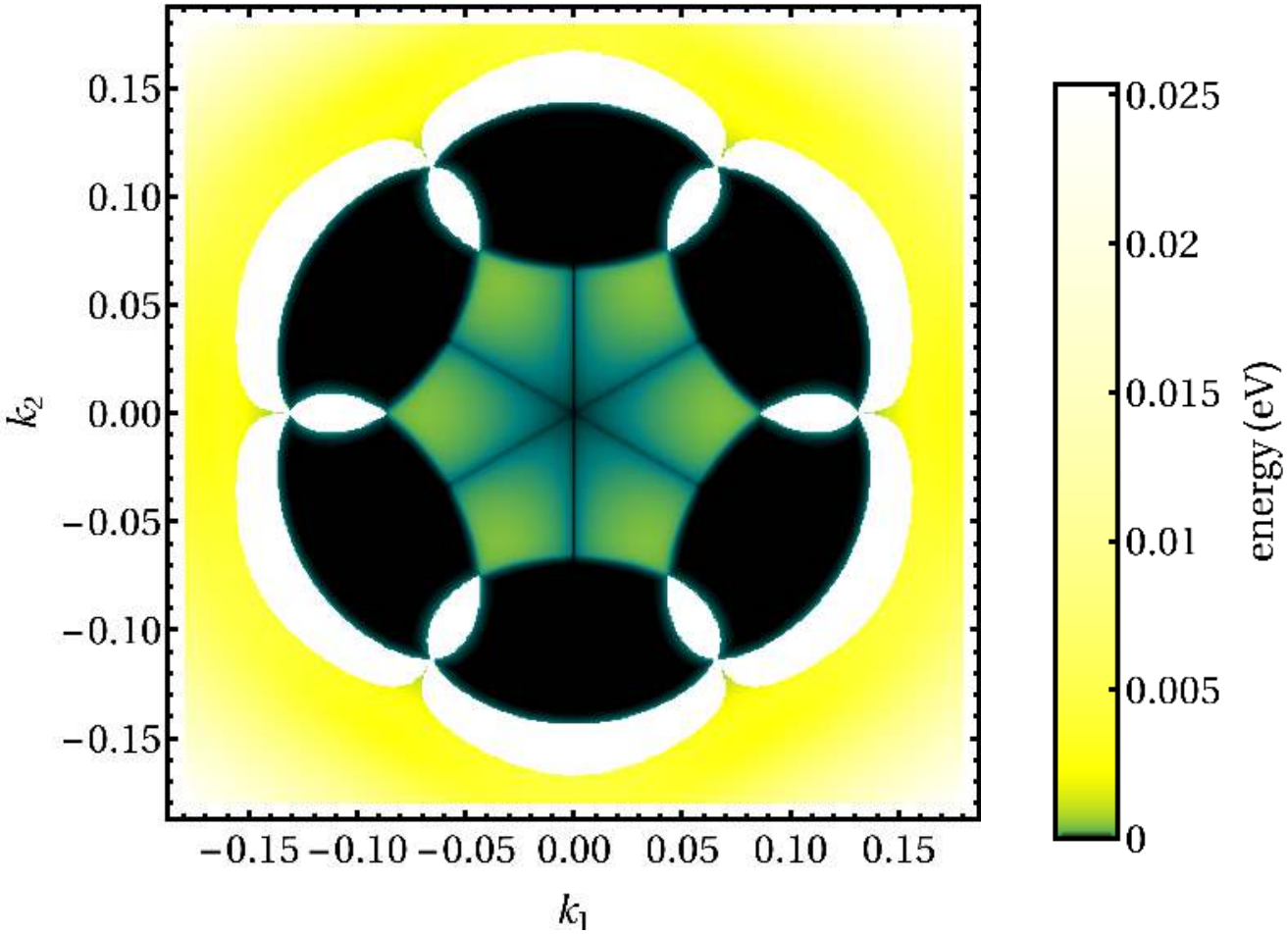}\\[2ex]%
\raisebox{0.5ex}{(b)}\includegraphics[width=0.95\columnwidth]{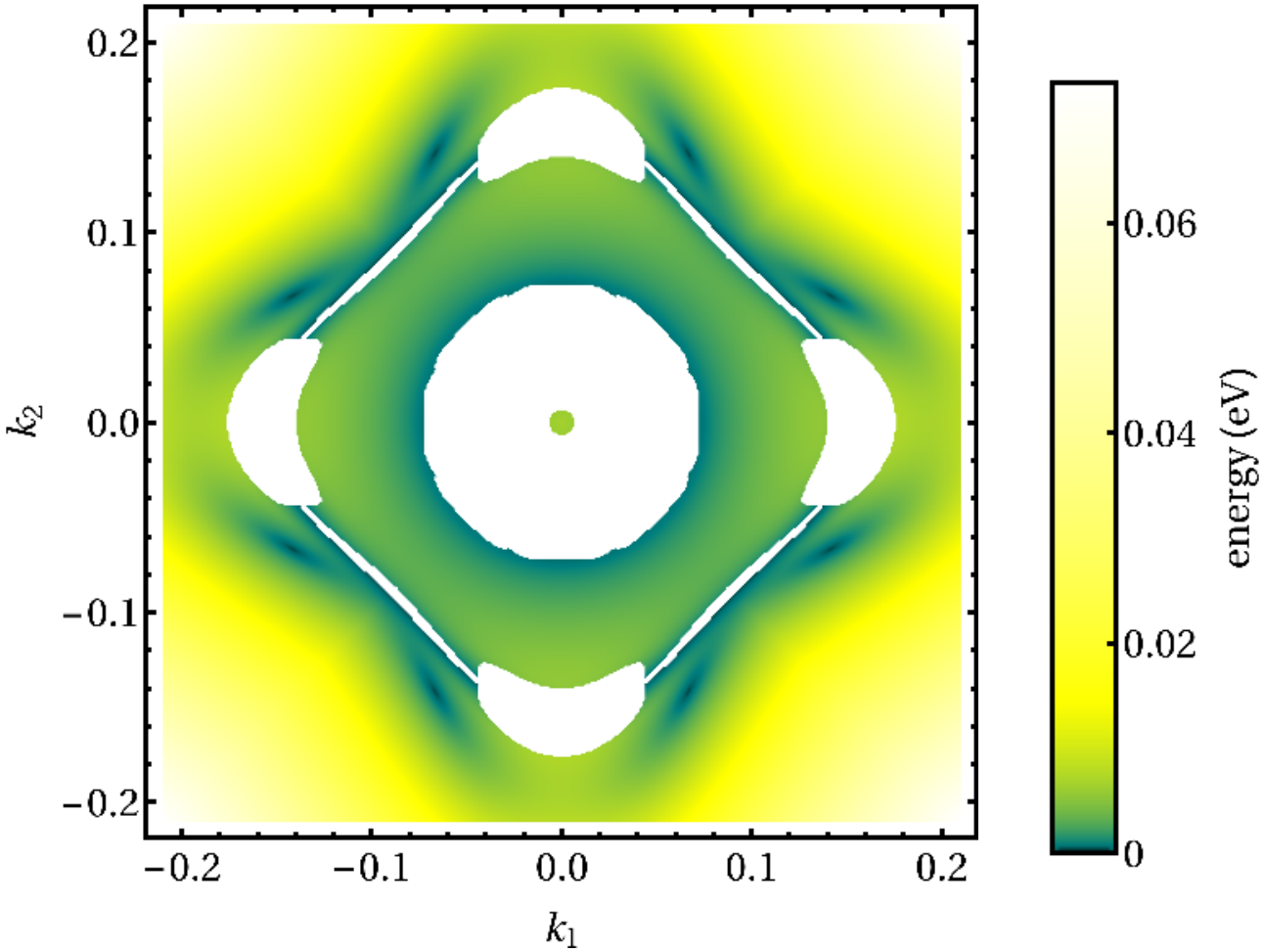}
\end{center}
\caption{Dispersion of surface states of YPtBi in the superconducting $A_1$ state for (a) the (111) surface and (b) the (100) surface, for a thickness of $W=4000$. At each momentum, the states occur in pairs with energies differing in their sign. The positive energy is plotted. Black regions correspond to flat surface bands. In the white regions, the states closest to the Fermi energy are bulk like.}
\label{fig.A1.surface}
\end{figure}

Figure \ref{fig.A1.surface} shows the dispersion of surface states at
the (111) and (100) surfaces. Outside of the projection of
the normal-state Fermi surface, we find weakly modified descendants of
the normal surface states shown in Fig.\ \ref{fig.normal.surface}.
Figure \ref{fig.A1.surface}(b) shows that the surface bands are mostly
gapped out by superconducting pairing. However, there are eight
symmetry-related points where the gap closes.
For the (111) surface, Fig.\ \ref{fig.A1.surface}(a), the projections
of the six nodal rings are clearly visible as overlapping
ellipses. For $\mathbf{k}_\|$ points within the projection of a single
nodal ring, we observe nondegenerate flat bands. These are reminiscent
of the flat bands predicted for spin-$1/2$ noncentrosymmetric nodal
superconductors \cite{TMY10,STY11,BST11,ScR11,SBT12,ScB15,BTS13,STB13}.
They indeed have the same origin: The model Hamiltonian is invariant under
time reversal and charge conjugation and thus also under their
product, i.e., chiral symmetry, which acts as 
\begin{equation}
S\, \mathcal{H}(\mathbf{k})\, S^\dagger = - \mathcal{H}(\mathbf{k}) ,
\end{equation}
where the unitary matrix $S$ reads
\begin{equation}
S = \left(\begin{array}{cc}
    0 & U_T \\
    U_T & 0
  \end{array}\right) .
\end{equation}
Let $U_S$ be a unitary matrix that diagonalizes the chiral operator $S$ so that
\begin{equation}
U_S\, S\, U_S^\dagger = \left(\begin{array}{cc}
  i\mathbb{1}_4 & 0 \\
  0 & -i\mathbb{1}_4
  \end{array}\right) ,
\end{equation}
where $\mathbb{1}_4$ is the $4\times 4$ identity matrix. $U_S$ transforms the Hamiltonian into block-off-diagonal form,
\begin{equation}
U_S\, \mathcal{H}(\mathbf{k})\, U_S^\dagger = \left(\begin{array}{cc}
  0 & D(\mathbf{k}) \\
  D^\dagger(\mathbf{k}) & 0
  \end{array}\right) .
\end{equation}
For any $\mathbf{k}_\|$ for which $\det D(\mathbf{k})\neq 0$ for all $k_\perp$, i.e., any $\mathbf{k}_\|$ not on a projected node, we can define the winding number
\begin{equation}
W(\mathbf{k}_\|) = \frac{1}{2\pi}\, \mathrm{Im} \int dk_\perp\,
  \frac{\partial}{\partial k_\perp}\, \ln \det D(\mathbf{k}) ,
\label{winding.2}
\end{equation}
where the integral is over the direction perpendicular to the $\mathbf{k}_\|$ plane and
\begin{equation}
\mathbf{k} = \mathbf{k}_\| + k_\perp\, \frac{1}{\sqrt{3}} \left(\begin{array}{@{}c@{}}
  1 \\ 1 \\ 1
  \end{array}\right) .
\end{equation}
Continuity of the function $\mathbf{k}\mapsto D(\mathbf{k})$ and the
properties of the branch point of the logarithm imply that
$W(\mathbf{k}_\|)\in \mathbb{Z}$. For our model, $W(\mathbf{k}_\|)=\pm
1$ for $\mathbf{k}_\|$ within the projection of a single nodal ring
and $W(\mathbf{k}_\|)=0$ otherwise. We indeed find nondegenerate flat
bands in exactly these regions.
In the limit of infinite thickness, these surface states are at zero energy.
They then formally become twofold degenerate but this is just due to the
double counting of states in the Nambu formalism. One can interpret them as a
pair of Majorana states for each $\mathbf{k}_\|$, which are localized
at opposite surfaces. In the regions where the projected nodal rings
overlap, there are no flat-band surface states. Here, winding numbers
$+1$ and $-1$ from the two rings add up to zero and the effective
one-dimensional system is trivial. 

For the (100) surface, Fig.\ \ref{fig.A1.surface}(b), four of the
nodal rings are viewed edge on and therefore do not lead to flat
surface bands. The other two are projected on top of each other.
The winding numbers $\pm 1$ from these two nodal rings add up
to zero, and so the argument used for the (111) surface does not
predict zero-energy flat bands. In agreement with this, we
do not find flat bands in this region and in fact also no dispersive
surface bands. The small green circle in the center of Fig.\
\ref{fig.A1.surface}(b) is likely a finite-size effect; the
spectrum here does not show a split-off band.

Furthermore, there are dispersive surface bands for small
$\mathbf{k}_\|$, where the normal-state Fermi surface has been gapped
out. For the (111) surface, Fig.\ \ref{fig.A1.surface}(a), their
energy goes to zero along lines (``arcs'') connecting the $\Gamma$
point with the flat bands. These arcs are \emph{not} of the same
origin as the ones predicted for noncentrosymmetric superconductors
with $C_{4v}$ symmetry in Refs.\ \cite{BST11,SBT12}, which result from
a $\mathbb{Z}_2$ invariant protected by an additional
time-reversal-like symmetry in two-dimensional planes in momentum
space. Instead, they rely on the presence of mirror symmetries.  
Of the six mirror planes in $T_d$, the three defined by the equations
$k_x-k_y=0$, $k_y-k_z=0$, and $k_z-k_x=0$ are perpendicular to the
(111) plane, i.e., the $\mathbf{k}_\|$ plane, and thus their
projections are straight lines. We observe that the arcs form parts of
these lines. For example, the mirror symmetry with respect to the
($1\bar{1}0$) plane with $k_x-k_y=0$ is expressed as 
\begin{equation}
M\, \mathcal{H}(k_y,k_x,k_z)\, M^\dagger = \mathcal{H}(k_x,k_y,k_z) ,
\label{mirror.2}
\end{equation}
with the unitary matrix
\begin{align}
M &= \left(\begin{array}{cc}
  e^{-i \pi (J_x - J_y)/\sqrt{2}} & 0 \\
  0 & (e^{-i \pi (J_x - J_y)/\sqrt{2}})^*
  \end{array}\right) \nonumber \\
&= \left(\begin{array}{cc}
  e^{-i \pi (J_x - J_y)/\sqrt{2}} & 0 \\
  0 & e^{i \pi (J_x + J_y)/\sqrt{2}}
  \end{array}\right) .
\end{align}
The ($1\bar{1}0$) plane corresponds to $k_1=0$, i.e., the $k_2$ axis in Fig.\ \ref{fig.A1.surface}(a), while $k_2$ and $k_\perp$ do not change under this reflection. Hence, we can write
\begin{equation}
M\, \mathcal{H}(k_1=0,k_2,k_\perp)\, M^\dagger = \mathcal{H}(k_1=0,k_2,k_\perp) .
\label{mirror.4}
\end{equation}
This is now a symmetry at a single $\mathbf{k}$ point, not one connecting two points. Let $U_M$ be a unitary matrix that diagonalizes $M$ in such a way that
\begin{equation}
U_M\, M\, U_M^\dagger = \left(\begin{array}{cc}
  i \mathbb{1}_4 & 0 \\
  0 & -i \mathbb{1}_4
  \end{array}\right) .
\end{equation}
Applying this transformation to the Hamiltonian on the mirror plane makes it block diagonal,
\begin{equation}
U_M\, \mathcal{H}(0,k_2,k_\perp)\, U_M^\dagger
  = \left(\begin{array}{cc}
    \mathcal{H}_+(k_2,k_\perp) & 0 \\
    0 & \mathcal{H}_-(k_2,k_\perp)
  \end{array}\right) ,
\end{equation}
where $\mathcal{H}_+(k_2,k_\perp)$ [$\mathcal{H}_-(k_2,k_\perp)$] is the Bogoliubov-de Gennes Hamiltonian in the sector with mirror eigenvalue $i$ ($-i$). The chiral operator is also invariant under reflection and is block diagonalized by the same transformation,
\begin{equation}
U_M\, S\, U_M^\dagger =  \left(\begin{array}{cc}
  \tilde S & 0 \\
  0 & -\tilde S
  \end{array}\right) .
\end{equation}
Note that the diagonal blocks are identical up to an irrelevant sign. The blocks $\mathcal{H}_\pm(k_2,k_\perp)$ possess chiral symmetry with respect to $\tilde S$:
\begin{equation}
\tilde S\, \mathcal{H}_\pm(k_2,k_\perp)\, \tilde S^\dagger = - \mathcal{H}_\pm(k_2,k_\perp) .
\end{equation}
With the $4\times 4$ unitary matrix $U_{\tilde S}$ that diagonalizes $\tilde S$, we can now separately transform the Hamiltonians $\mathcal{H}_\pm(k_2,k_\perp)$ in the two mirror-parity sectors into block-off-diagonal form,
\begin{equation}
U_{\tilde S}\, \mathcal{H}_\pm(k_2,k_\perp)\, U_{\tilde S}^\dagger = \left(\begin{array}{cc}
  0 & D_\pm(k_2,k_\perp) \\
  D_\pm^\dagger(k_2,k_\perp) & 0
  \end{array}\right) .
\end{equation}
The winding number in Eq.\ (\ref{winding.2}) can also be written as
\begin{equation}
W(\mathbf{k}_\|) = -\frac{1}{4\pi i} \int dk_\perp \, \Tr S\, \mathcal{H}(\mathbf{k})^{-1}
  \frac{\partial}{\partial k_\perp}\, \mathcal{H}(\mathbf{k}) .
\end{equation}
We now rewrite this winding number on the mirror plane, suppressing the arguments $(k_2,k_\perp)$ of $\mathcal{H}_\pm$,
\begin{align}
W(k_2) &= -\frac{1}{4\pi i} \int dk_\perp\, \Tr \bigg\{
  \left(\begin{array}{cc}
    \tilde S \mathcal{H}_+^{-1} & 0 \\
    0 & -\tilde S \mathcal{H}_-^{-1}
  \end{array}\right) \nonumber \\
&\quad{}\times \frac{\partial}{\partial k_\perp}\,
  \left(\begin{array}{cc}
    \mathcal{H}_+ & 0 \\
    0 & \mathcal{H}_-
  \end{array}\right) \bigg\} \nonumber \\
&= W_+(k_2) - W_-(k_2) ,
\end{align}
with
\begin{align}
W_\pm(k_2) &\equiv -\frac{1}{4\pi i} \int dk_\perp\, \Tr \tilde S\,
  \mathcal{H}_\pm^{-1}\, \frac{\partial}{\partial k_\perp}\,
  \mathcal{H}_\pm \nonumber \\
&= \frac{1}{2\pi}\, \mathrm{Im} \int dk_\perp\,
  \frac{\partial}{\partial k_\perp}\, \ln \det D_\pm(k_2,k_\perp) .
\end{align}
We find that $W_+(k_2)=W_-(k_2)=-1$ on the arcs. Here, the normal winding number is $W(\mathbf{k}_\|)=-1+1=0$ but the two nontrivial mirror winding numbers $W_\pm(k_2)=-1$ leads to two zero-energy states. The arcs are indeed twofold degenerate if the double counting introduced by the Nambu formalism is corrected for: two pairs of helical Majorana bands cross at $k_1=0$. The splitting between the Majorana bands is due to the ASOC.
In the flat-band regions, one of $W_\pm(k_2)$ equals $-1$ and the other vanishes, leading to $W(\mathbf{k}_\|)=\pm 1$, as discussed above. For larger $k_2$, outside of the flat-band regions, we find $W_+(k_2)=W_-(k_2)=0$ and there is no arc.

The surface states are nondegenerate, except at the arcs. In particular, the flat bands are nondegenerate, as noted above. Therefore, the states are spin polarized and the absolute value of the spin polarization is $1/2$. Physically, the spin polarization of states results from the ASOC. For the single-band case and various point groups, the spin polarization has been calculated in Ref.\ \cite{BST15}. In the present case, we have to distinguish between the effective spin $\mathbf{J}$ of length $3/2$ and the electronic spin $\mathbf{S}$ of length $1/2$. In the half-Heusler compounds, the total angular momentum $\mathbf{J}$ results from the combination of the electronic spin $\mathbf{S}$ with an orbital angular momentum $\mathbf{L}$ of length $1$. The spin operators are then obtained by projecting $S_\nu \otimes \mathbb{1}_3$, $\nu=x,y,z$ onto the subspace of total angular momentum $3/2$ \cite{AJB01,JSM06}. This simply gives $\mathbf{S} = \mathbf{J}/3$ \cite{endnote.orbitalspin}.

\begin{figure}[tb]
\begin{center}
\includegraphics[width=0.95\columnwidth]{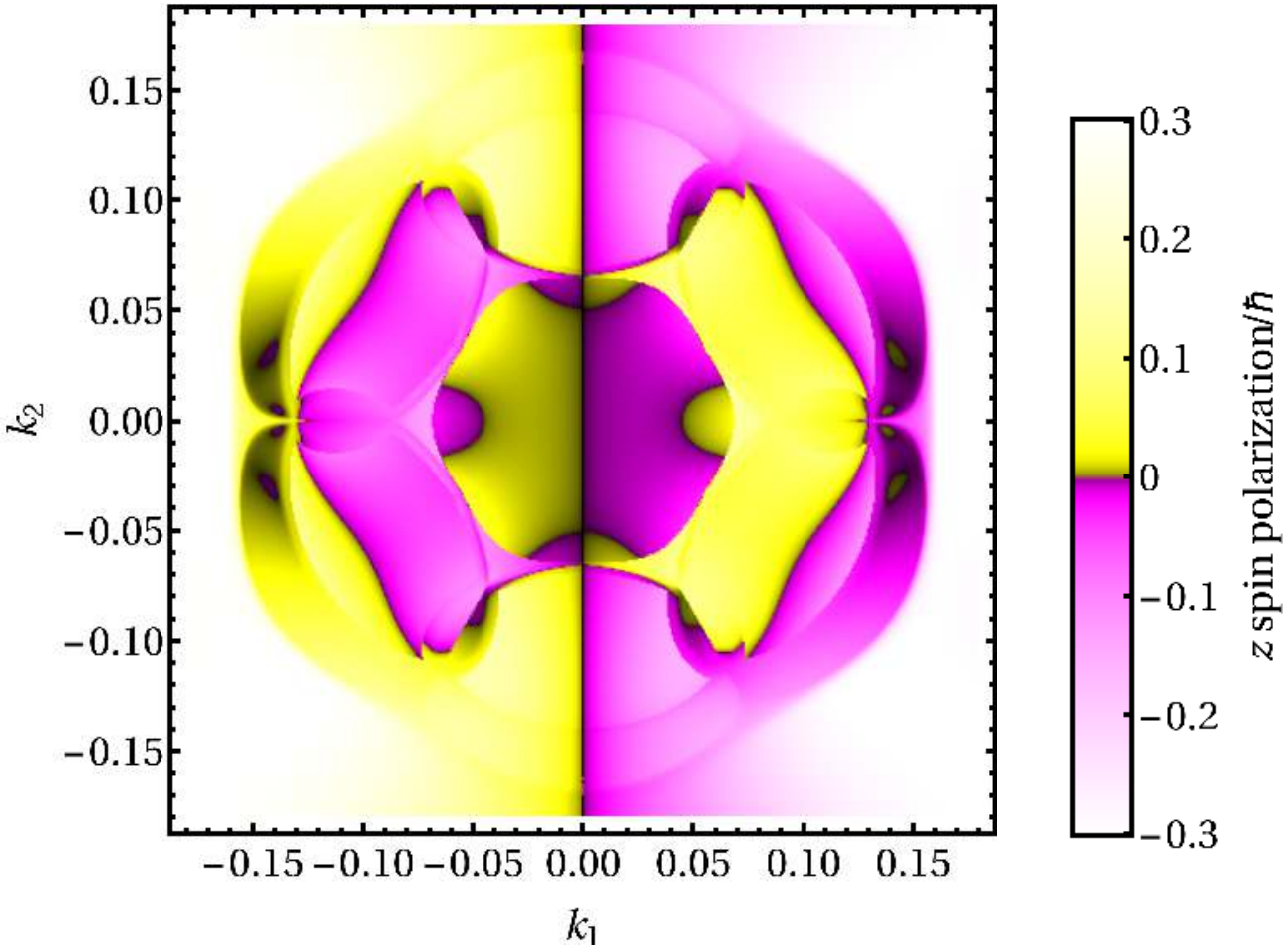}
\end{center}
\caption{Spin polarization in the \textit{z} direction, $\langle S_z\rangle = \langle J_z/3\rangle$ versus surface momentum $\mathbf{k}_\|$ of the lowest-energy state for a (111) slab with $W=4000$ of YPtBi in the $A_1$ state. At each momentum, the spectrum is symmetric. For states of nonzero energy, the negative-energy state is used, whereas for degenerate zero-energy states, the superposition localized at the $l=0$ surface is used.}
\label{fig.A1.spin}
\end{figure}

We show the $z$ component of the spin polarization $\langle\mathbf{S}\rangle$ at the (111) surface in Fig.\ \ref{fig.A1.spin}. 
Making use of the threefold rotation symmetry of the slab with respect to the (111) direction, which is perpendicular to the $\mathbf{k}_\|=(k_1,k_2)$ plane, the $x$ and $y$ components can be obtained by rotating the plot by $\pm 120^\circ$ (not shown).
Moreover, the mirror planes perpendicular to the surface are visible in the plot. For example, the mirror symmetry of the Bogoliubov-de Gennes Hamiltonian with respect to the ($1\bar{1}0$) plane is expressed by Eq.\ (\ref{mirror.2}). In the mirror plane that intersects the $\mathbf{k}_\|$ plane along the $k_2$ axis this relation turns into the invariance (\ref{mirror.4}) at fixed momentum. The same transformation also leaves the spin component $(J_x-J_y)/\sqrt{2}$ invariant. The surface states must therefore be eigenstates of $(J_x-J_y)/\sqrt{2}$. Noting that the components $J_z$ and $(J_x+J_y)/\sqrt{2}$ are orthogonal to $(J_x-J_y)/\sqrt{2}$, it is elementary to show that the operators  $J_z$ and $(J_x+J_y)/\sqrt{2}$ have vanishing diagonal matrix elements in an eigenbasis of $(J_x-J_y)/\sqrt{2}$ so that $\langle J_z\rangle=0$ and $\langle J_x\rangle = -\langle J_y\rangle$.


\subsection{$T_2$ pairing: inflated bulk nodes and Chern Fermi arcs}
\label{sec.T2}

For the half-Heusler compounds, time-re\-ver\-sal-sym\-me\-try-brea\-king
states are realized by forming linear combinations of the pairing matrices
$\Gamma_r$ in Eqs.\ (\ref{Gamma.E.1})--(\ref{Gamma.T2.3}) belonging to the
same irrep, with complex coefficients. We here focus on the pairing states
belonging to the three-dimensional irrep $T_2$. They are characterized by
the gap matrix
\begin{equation}
\Delta_{ij} = 2\delta_{ij}\, \Delta^0_{T_2}\,
  (l_1 \Gamma_{T_2,1} + l_2 \Gamma_{T_2,2} + l_3 \Gamma_{T_2,3})
\end{equation}
in terms of the complex three-component order parameter $\mathbf{l} = (l_1,l_2,l_3)$. A free-energy expansion \cite{SU91} shows that the possible equilibrium states are $\mathbf{l}=(1,0,0)$, $(1,1,1)$, $(1,e^{2\pi i/3},e^{4\pi i/3})$, $(1,i,0)$, and states related to these by point-group operations. The third and fourth state in the list break time-reversal symmetry because of the complex phase factors. For weak pairing, one of these states has the lowest free energy of all $T_2$ pairing states \cite{BWW16}. As a representative of pairing that breaks time-reversal symmetry, we examine the state with $\mathbf{l}=(1,i,0)$, i.e.,
\begin{equation}
\Delta_{ij} = 2 \delta_{ij}\, \Delta^0_{T_2}\, (\Gamma_{T_2,1} + i\,\Gamma_{T_2,2}) .
\label{Deltaij.T2.2}
\end{equation}

\subsubsection{Bulk nodal structure}

In the centrosymmetric limit, where the ASOC vanishes, and for
infinitesimal pairing amplitude $\Delta^0_{T_2}$,
the superconducting gap has both point and line nodes \cite{BWW16}.
The point nodes are located at the intersections of the $k_z$ axis with the
Fermi surfaces, which have first-order touching points at these intersections.
The line nodes exist at the intersections of the $k_xk_y$ plane with both
Fermi surfaces. The same nodal structure is found also for stronger pairing
if the pairing is purely intraband.
Point nodes are not surprising since the superconductor belongs to class D \cite{Zir96}, for which point nodes away from high-symmetry points are protected by a $\mathbb{Z}$ topological invariant \cite{ChS14,ScB15,CTS16}. The invariant is a first Chern number, which for our model evaluates to $-2$ ($+2$) for the point node on the positive (negative) $k_z$ axis \cite{endnote.Chern}.
However, class D does not yield a topological invariant for line nodes in high-symmetry planes. These rely on a lattice symmetry, namely on the twofold rotation axis along $\hat{\mathbf{z}}$, which acts as
\begin{equation}
\tilde{\mathcal{P}}\, \mathcal{H}(-k_x,-k_y,k_z)\, \tilde{\mathcal{P}}^\dagger
  = \mathcal{H}(k_x,k_y,k_z) ,
\end{equation}
with
\begin{equation}
\tilde{\mathcal{P}} = \left(\begin{array}{cc}
  ie^{-i \pi J_z} & 0 \\
  0 & -ie^{i \pi J_z}
  \end{array}\right) .
\end{equation}
In the $k_xk_y$ plane, the twofold rotation maps $(k_x,k_y,0)$ onto $(-k_x,-k_y,0)$ and hence acts like spatial inversion. The product of charge conjugation $\mathcal{C}$ and the pseudo-inversion $\tilde{\mathcal{P}}$ maps $(k_x,k_y,0)$ onto itself and causes the spectrum to be symmetric for each momentum in the $k_xk_y$ plane.
We further find that this product squares to $(\mathcal{C}\tilde{\mathcal{P}})^2=+1$ since the antiunitary charge-conjugation operator reads $\mathcal{C}=U_C\mathcal{K}$ with $U_C = \tau_1 \otimes \mathbb{1}_4$, where $\tau_1$ is a Pauli matrix in particle-hole space. Such a symmetry ensures that nodes of codimension $1$ can have a $\mathbb{Z}_2$ topological invariant \cite{KST14,ZSW16}. In the two-dimensional $k_xk_y$ plane, line nodes can thus be topologically stable in the presence of a twofold rotation axis perpendicular to the plane. The method of Ref.\ \cite{ABT17} can be applied to construct this invariant in terms of a Pfaffian of $\mathcal{H}(k_x,k_y,0)$.

The $T_2$ state of Eq.\ (\ref{Deltaij.T2.2}) is of particular relevance since there is experimental evidence for line nodes in YPtBi \cite{KWN16}. The other symmetry-allowed and energetically favorable $E$ and $T_2$ pairing states do not have symmetry-protected line nodes for vanishing ASOC \cite{BWW16}.

For the centrosymmetric variant with point group $O_h$, for which the
ASOC is forbidden by symmetry, nodes of the superconducting gap are
generically \emph{inflated} into two-dimensional Bogoliubov Fermi
surfaces for multiband pairing \cite{ABT17,BzS17}. In centrosymmetric
multiband superconductors that spontaneously break time-reversal
symmetry but satisfy $\mathcal{CP}$ symmetry (the product of charge
conjugation and inversion) squaring to $(\mathcal{CP})^2=+1$, nodal
points and nodal lines are replaced by spheroidal and toroidal
Bogoliubov Fermi surfaces, respectively. These Fermi surfaces are
protected by a $\mathbb{Z}_2$ topogical invariant, which can be
expressed in terms of a Pfaffian \cite{KST14,ZSW16,ABT17,BzS17}. These
results do not carry over to the present case since the $T_d$ point
group is not centrosymmetric and thus inversion and $\mathcal{CP}$
symmetry are absent. The $\mathbb{Z}_2$ number \cite{ABT17,BzS17}
cannot even be defined. Nevertheless, Bogoliubov Fermi surfaces can exist:
Volovik \cite{Vol89} has pointed out that  Fermi surfaces can appear
in multiband systems if both inversion and time-reversal symmetry
are broken and the interband pairing potential is sufficiently
large. Examples of this are realized in Fulde-Ferrell \cite{FuF64}
pairing states in single-band systems with Rashba spin-orbit
coupling in an applied magnetic field~\cite{XZZ15}, as well as in
the proposed \cite{Var97} coexistence state of \textit{d}-wave
superconductivity and loop-current order in $\mathrm{YBa_2Cu_3O}_{7-\delta}$~\cite{WaV13}.
Note that these proposals require both the inversion and the time-reversal-symmetry breaking to be extrinsic to the superconducting state; in contrast, in our system the time-reversal-symmetry breaking is intrinsic to the superconductivity.

The presence of Fermi surfaces in these systems can be understood as
follows: due to the absence of $\mathcal{CP}$ and $\mathcal{CT}$
symmetries, the spectrum at fixed momentum $\mathbf{k}$ is not
symmetric. Hence, band crossings or avoided crossings generically do
not occur at the Fermi energy and symmetry thus does not dictate any
gap opening there. It is thus possible for bands to cross the Fermi
energy. Since the band energies are continuous functions of momentum,
the crossings are generically two-dimensional Fermi surfaces. Of
course, one expects that superconductivity is only energetically
favorable if gaps do open at the Fermi energy, but this need not
happen everywhere on the normal-state Fermi surface.

The nodal structure both for infinitesimal and finite pairing is best analyzed in terms of the determinant of the Bogoliubov-de Gennes Hamiltonian $\mathcal{H}(\mathbf{k})$ given in Eq.\ (\ref{H.super.3a}). The determinant is of course the product of the eigenenergies and thus its zeros coincide with the nodes.
The determinant is expanded in the pairing amplitude $\Delta^0_{T_2}$, which is assumed to be real,
\begin{equation}
\det\mathcal{H}(\mathbf{k}) = \det\mathcal{H}(\mathbf{k})\big|_{\Delta=0}
  + g_2(\mathbf{k}) (\Delta^0_{T_2})^2 + g_4(\mathbf{k}) (\Delta^0_{T_2})^4 .
\label{T2.detexp.2}
\end{equation}
This expansion is exact; higher orders do not occur since
\begin{equation}
\Delta(\mathbf{k}) = 4 \Delta^0_{T_2} \left(\begin{array}{cccc}
    0 & 0 & 0 & 0 \\
    0 & 0 & 0 & i \\
    0 & 0 & 0 & 0 \\
    0 & -i & 0 & 0
  \end{array}\right)
\end{equation}
so that the matrix $\mathcal{H}(\mathbf{k})$ only contains
$\Delta^0_{T_2}$ linearly in four components. The determinant must be
even in $\Delta^0_{T_2}$ due to invariance under global phase
changes. The zero-order term in Eq.\ (\ref{T2.detexp.2}) is
nonnegative since the spectrum is symmetric and contains an even
number of pairs of eigenvalues. Hence,
$\det\mathcal{H}(\mathbf{k})\big|_{\Delta=0}$ generically (not at the
touching points) has second-order zeros forming a two-dimensional
manifold, namely the Fermi surface.

The determinant $\det\mathcal{H}(\mathbf{k})$ of the full Hamiltonian
is positive sufficiently far from the normal-state Fermi surface since
superconductivity is then a small correction. A negative determinant
at some momentum $\mathbf{k}$ thus implies, using the continuity of
the function $\mathbf{k} \mapsto \det\mathcal{H}(\mathbf{k})$, the
existence of a Bogoliubov Fermi surface surrounding it. For small
$\Delta^0_{T_2}$, this can only happen on the normal-state Fermi
surface, where the zero-order term vanishes. To leading order, the
existence of Fermi surfaces is determined by the coefficient
$g_2(\mathbf{k}_F)$ on the Fermi surface. If $g_2(\mathbf{k}_F)>0$
then the normal-state Fermi surface is gapped out in the vicinity of
$\mathbf{k}_F$. For $g_2(\mathbf{k}_F)<0$, there must be a Bogoliubov
Fermi surface in the superconducting state. If $g_2(\mathbf{k}_F)=0$
then it is necessary to go to higher orders.

\begin{figure}[tb]
\begin{center}
\raisebox{1ex}{(a)}\includegraphics[width=0.7\columnwidth]{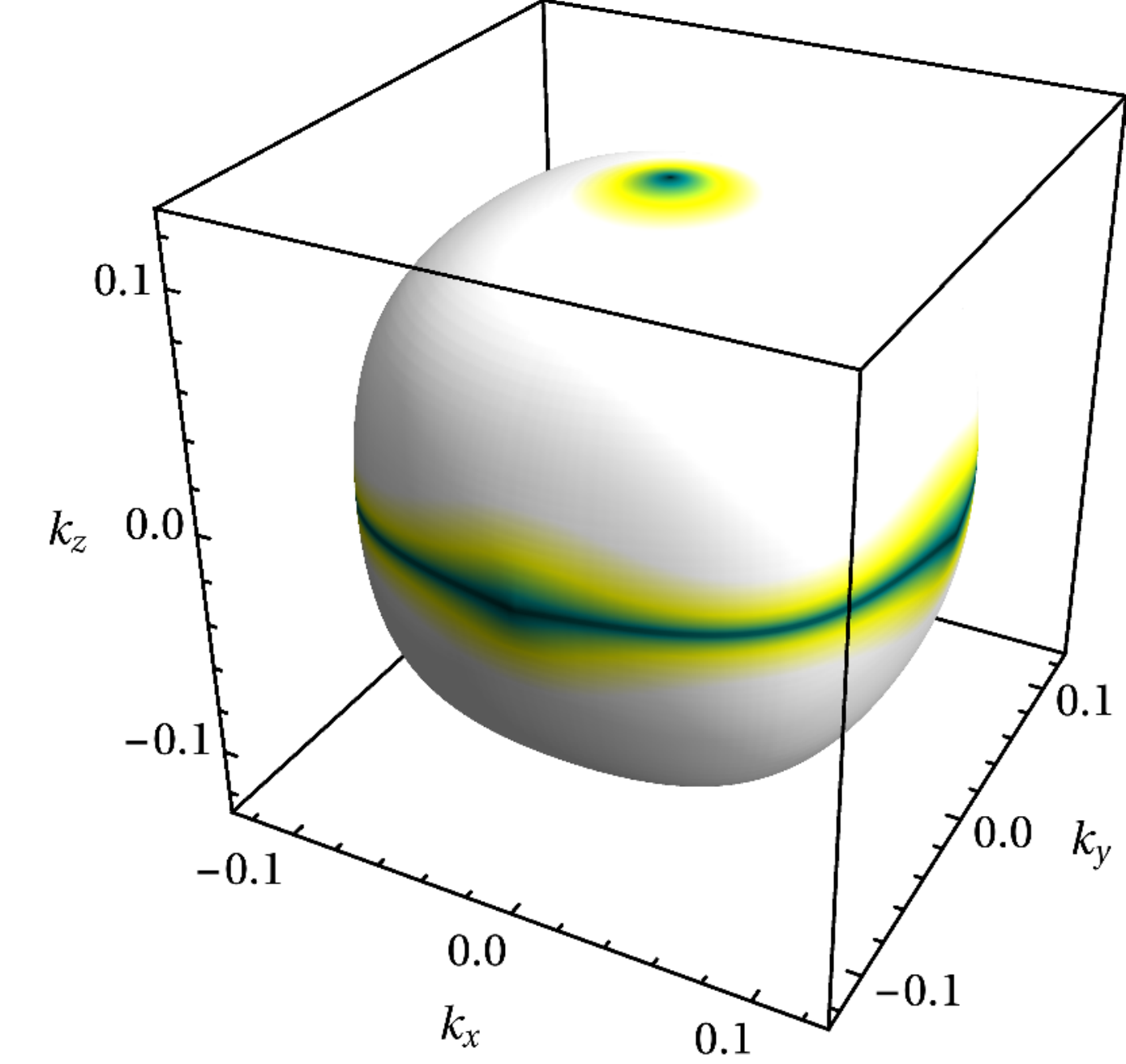}\\[2ex]%
\raisebox{1ex}{(b)}\includegraphics[width=0.7\columnwidth]{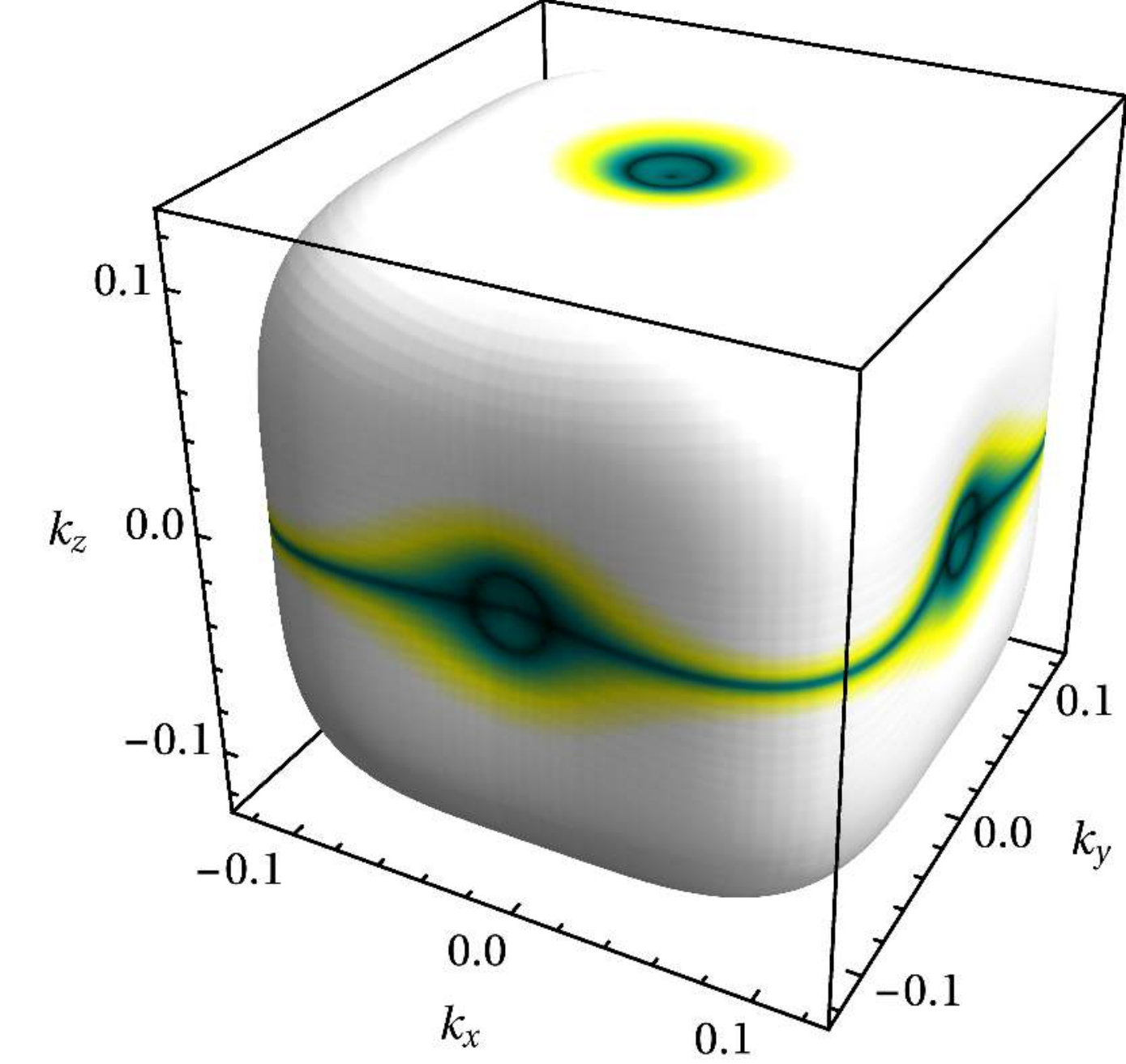}\\[2ex]%
\hspace{2em}\includegraphics[width=0.6\columnwidth]{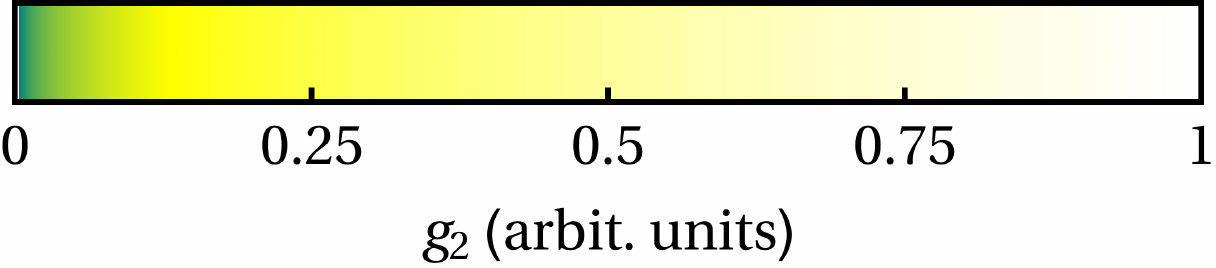}
\end{center}
\caption{Coefficient $g_2$ in the expansion of $\det\mathcal{H}(\mathbf{k})$ in the pairing amplitude, Eq.\ (\ref{T2.detexp.2}), for the $T_2$ pairing state with order parameter $\mathbf{l}=(1,i,0)$ on (a) the smaller and (b) the larger normal-state Fermi surface. The plots exhibit the nodal structure for infinitesimal pairing.}
\label{fig.T2.g2}
\end{figure}

It can be shown that $g_2(\mathbf{k}_F) \ge 0$ everywhere on the
normal-state Fermi surface. Moreover, $g_2$ vanishes where the $k_x$,
$k_y$, and $k_z$ axes intersect the Fermi surfaces. The proof of these
statements is given in the Appendix. Numerical results for $g_2$
on the Fermi surfaces are shown in Fig.\ \ref{fig.T2.g2}. The plots
show the nodal structure for infinitesimal pairing. Besides the zeros
on the coordinate axes found rigorously, $g_2$ also vanishes along the
equator of both surfaces and on additional nodal rings surrounding the
coordinate axes only on the larger surface. These new nodal rings are
reminiscent of the case of $A_1$ pairing in that they do not lie in
high-symmetry planes. However, unlike for $A_1$ pairing, their  size
is controlled by the ASOC. Everywhere else, the coefficient $g_2$ is
positive and the Fermi surface is gapped out in the superconducting
state. Hence, for infinitesimal pairing, the point and line nodes of
the single-band $T_2$ state survive but we obtain additional nodal
rings on one of the Fermi surfaces. 

\begin{figure}[tb]
\vspace{3ex}
\begin{center}
\raisebox{0.5ex}{(a)}\includegraphics[width=0.9\columnwidth]{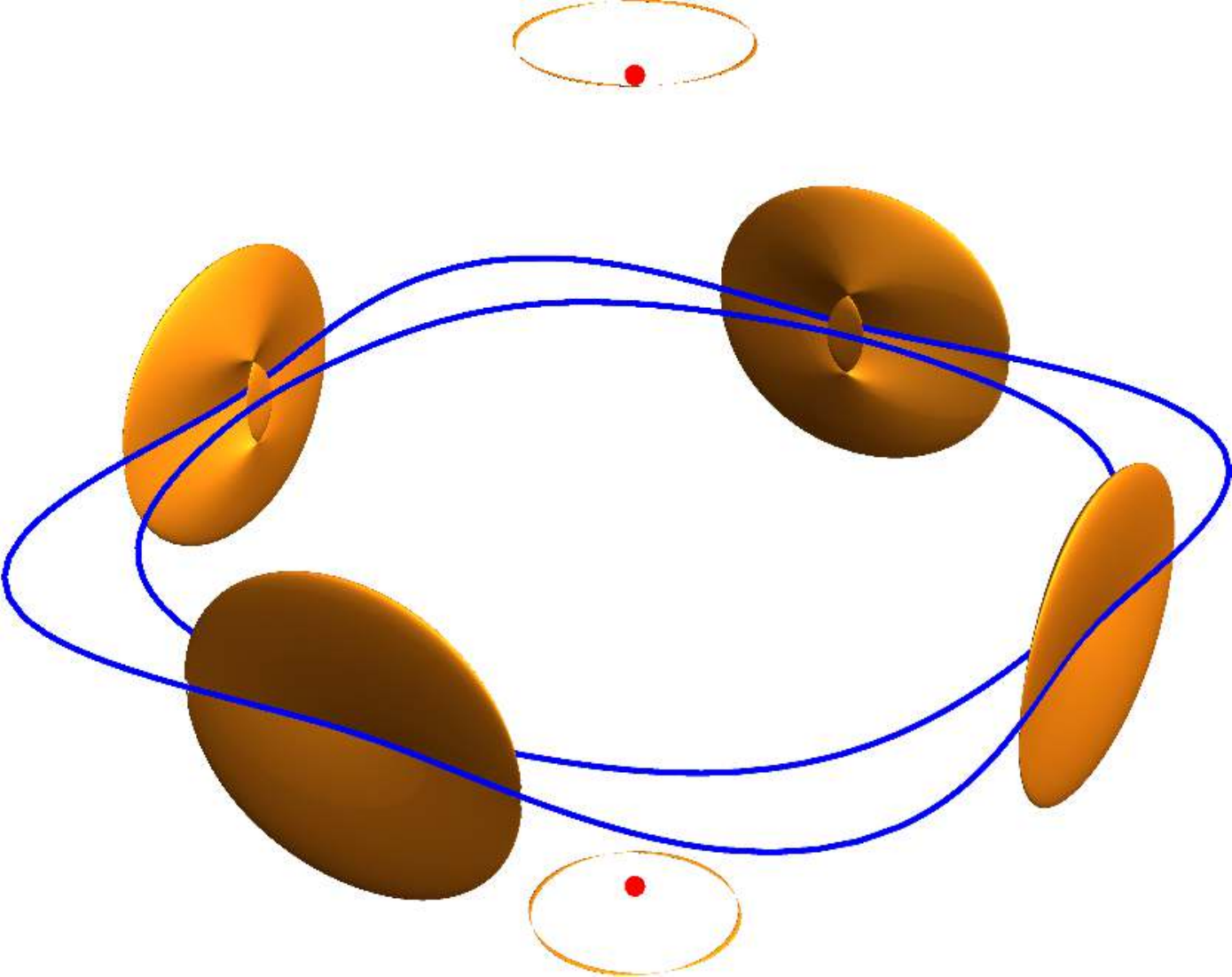}\\[2ex]%
\raisebox{0.5ex}{(b)}\includegraphics[width=0.55\columnwidth]{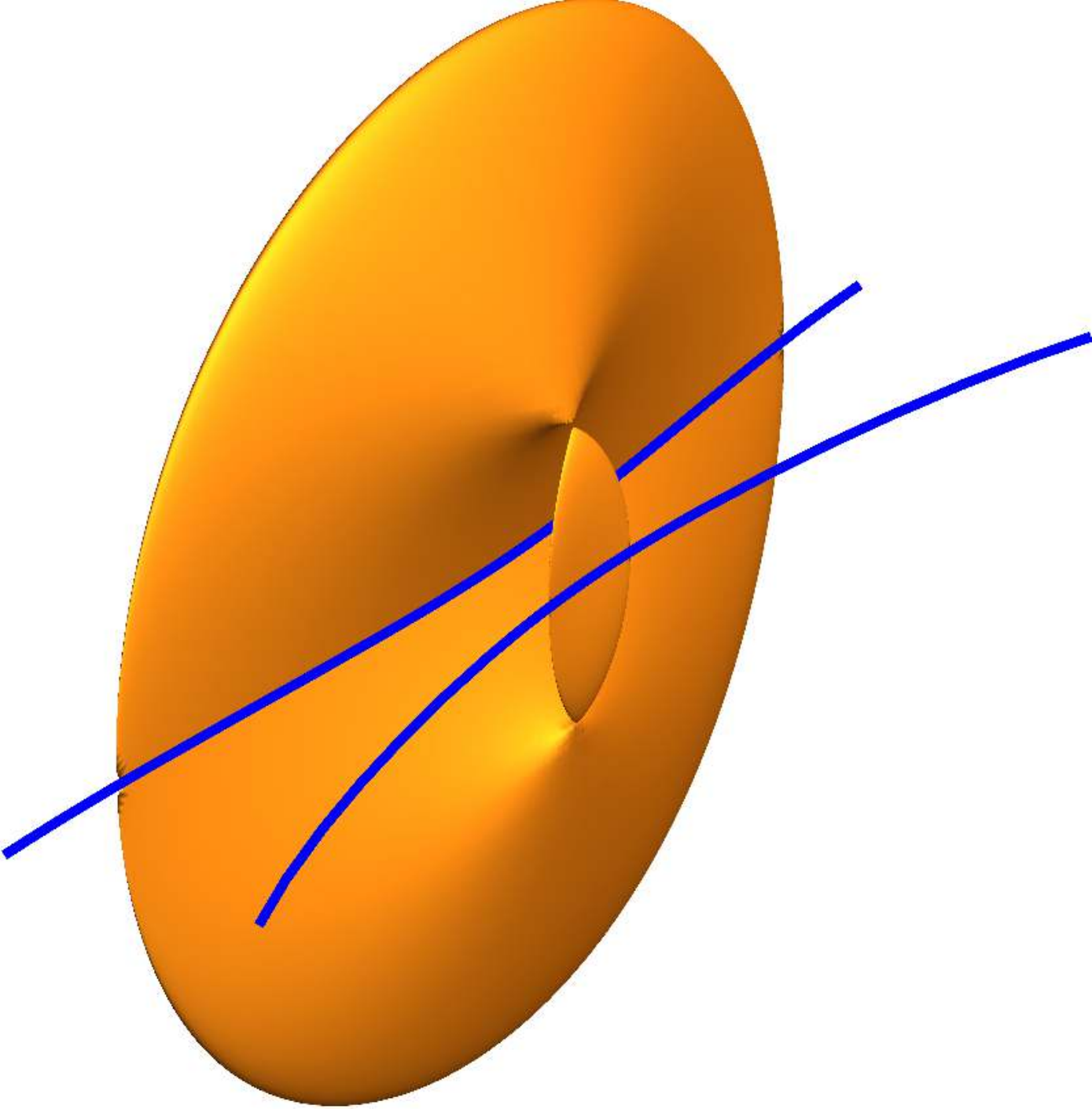}\\[2ex]%
\raisebox{0.5ex}{(c)}\hspace{1em}\includegraphics[width=0.55\columnwidth]{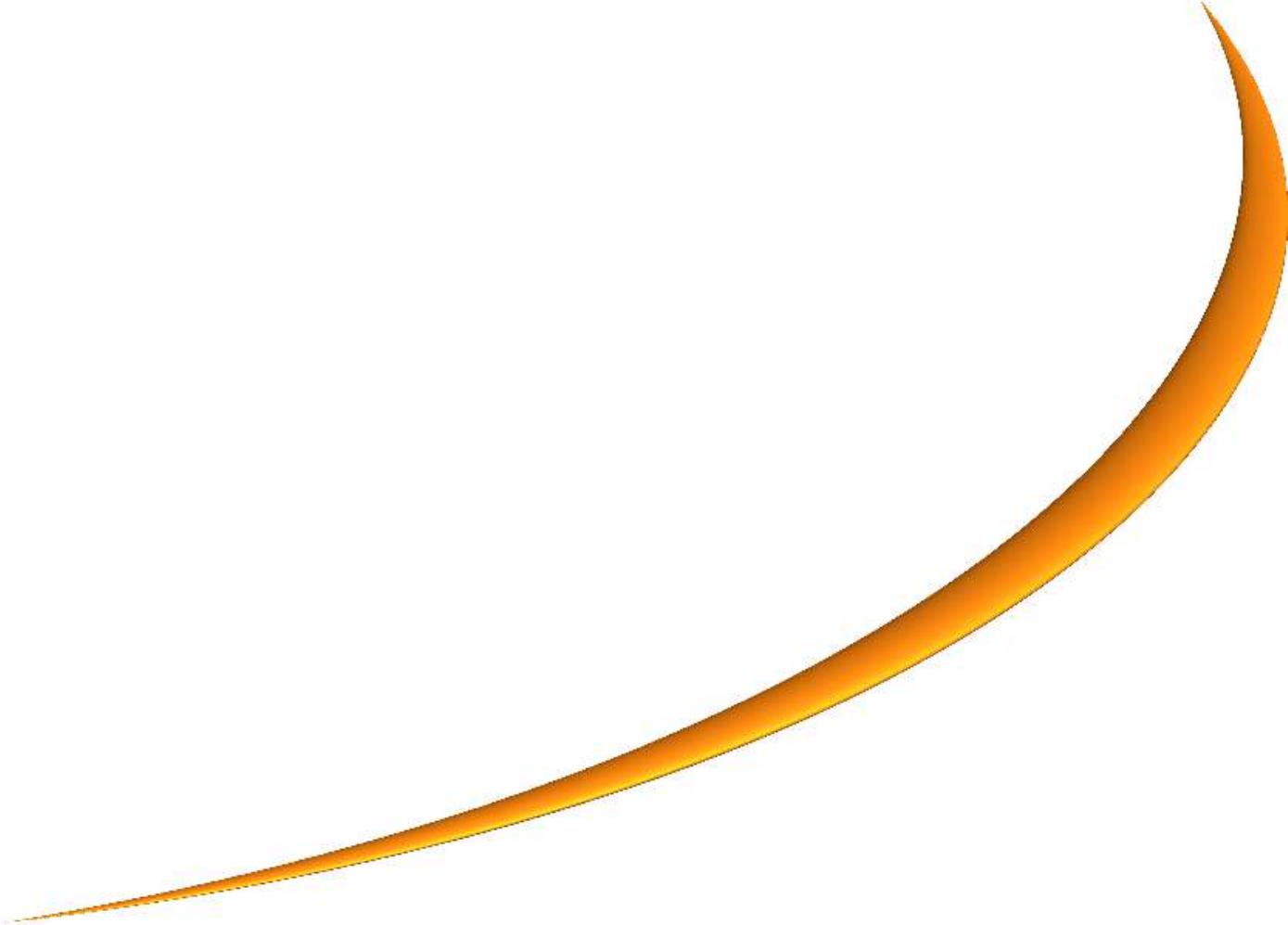}
\end{center}
\caption{(a) Nodes for the $T_2$ pairing state with order parameter $\mathbf{l}=(1,i,0)$, which breaks time-reversal symmetry, and pairing amplitude $\Delta^0_{T_2}=3\,\mathrm{meV}$. Red dots represent point nodes, blue lines line nodes, and orange surfaces Bogoliubov Fermi surfaces.
Momentum axes are omitted for clarity. Note, however, that the point nodes coincide with the ones in Fig.~\ref{fig.T2.g2}. (b) Enlargement of the system of Fermi pockets and line nodes close to the $k_x$ and $k_y$ axes. (c) Enlargement of one of the sickle-shaped Fermi pockets close to the point nodes.}
\label{fig.T2.nodes}
\end{figure}

If the pairing amplitude is not infinitesimal, we have to go beyond
second order in $\Delta^0_{T_2}$. However, using
Eq.\ (\ref{T2.detexp.2}) is inconvenient since the minima of
$\det\mathcal{H}(\mathbf{k})$ generically occur off the Fermi surface,
necessitating a double expansion in the pairing amplitude and the
deviation of $\mathbf{k}$ normal to the Fermi surface. Instead, we
have numerically examined $\det\mathcal{H}(\mathbf{k})$. The results
for $\Delta^0_{T_2}=3\,\mathrm{meV}$ are shown in
Fig.~\ref{fig.T2.nodes}.  

In more detail, we find the following results:
(\textit{i}) The two line nodes along the equators survive but the breaking of time-reversal symmetry splits their touching points on the $k_x$ and $k_y$ axes.
(\textit{ii}) In the vicinity of these points, regions with $\det\mathcal{H}(\mathbf{k})<0$ emerge, where the number of bands below (and above) the Fermi energy is odd. These regions are bounded by two-dimensional Bogoliubov Fermi surfaces (inflated nodes) of complex shape. Figure \ref{fig.T2.nodes}(b) shows an enlarged view of one set of these pockets, consisting of a large plate-like pocket and a smaller pocket that touches the larger one at two points. The surface of each of the two pockets crosses itself at one of the nodal lines. The pockets evolve from the additional nodal rings around the $k_x$ and $k_y$ axes in Fig.\ \ref{fig.T2.g2}(b), which are inflated for growing $\Delta^0_{T_2}$, and transform into the shapes in Fig.\ \ref{fig.T2.nodes}(b) through a series of Lifshitz transitions.
(\textit{iii}) There are still two point nodes on the $k_z$ axis. The quasiparticle dispersion close to these nodes is linear in the $k_z$ direction but quadratic in the two orthogonal directions. The point nodes have Chern numbers of $\pm 2$ \cite{endnote.Chern}.
(\textit{iv}) The nodal rings surrounding the $k_z$ axis on the larger Fermi surface, see Fig.\ \ref{fig.T2.g2}(b), are inflated into Bogoliubov Fermi surfaces with four pinch points for $|k_x|=|k_y|$, resulting in four sickle-shaped pockets for each ring. An enlargement of one of these pockets is shown in Fig.~\ref{fig.T2.nodes}(c).

It is worth emphasizing that point and line nodes coexist with
Bogoliubov Fermi pockets. As discussed above, point nodes protected
by an integer Chern number are allowed for class D, while line nodes
are protected by class D in conjunction with a twofold rotation symmetry.
Neither the multiband character nor the breaking of inversion symmetry by
the ASOC affect these symmetries.
The point nodes inherit the Chern numbers $\pm 2$ from the inversion-symmetric,
weak pairing limit. Since the model stays in class D when multiband pairing and ASOC
are switched on, these topological invariants could only vanish by merging in the
$k_xk_y$ plane. However, generically the point nodes can split into two each
with Chern numbers $\pm 1$. We find that this splitting is disallowed by
the combination of charge-conjugation
symmetry and fourfold rotoinversion symmetry about the \textit{z}-axis.
On the other hand, the Bogoliubov Fermi pockets are not
topologically protected. As noted above, bands crossing the
Fermi energy are allowed by the breaking of time-reversal symmetry, which
generically makes avoided band crossings happen away from the Fermi energy.

\begin{figure}[tb]
\begin{center}
\raisebox{1ex}{(a)}\includegraphics[width=0.95\columnwidth]{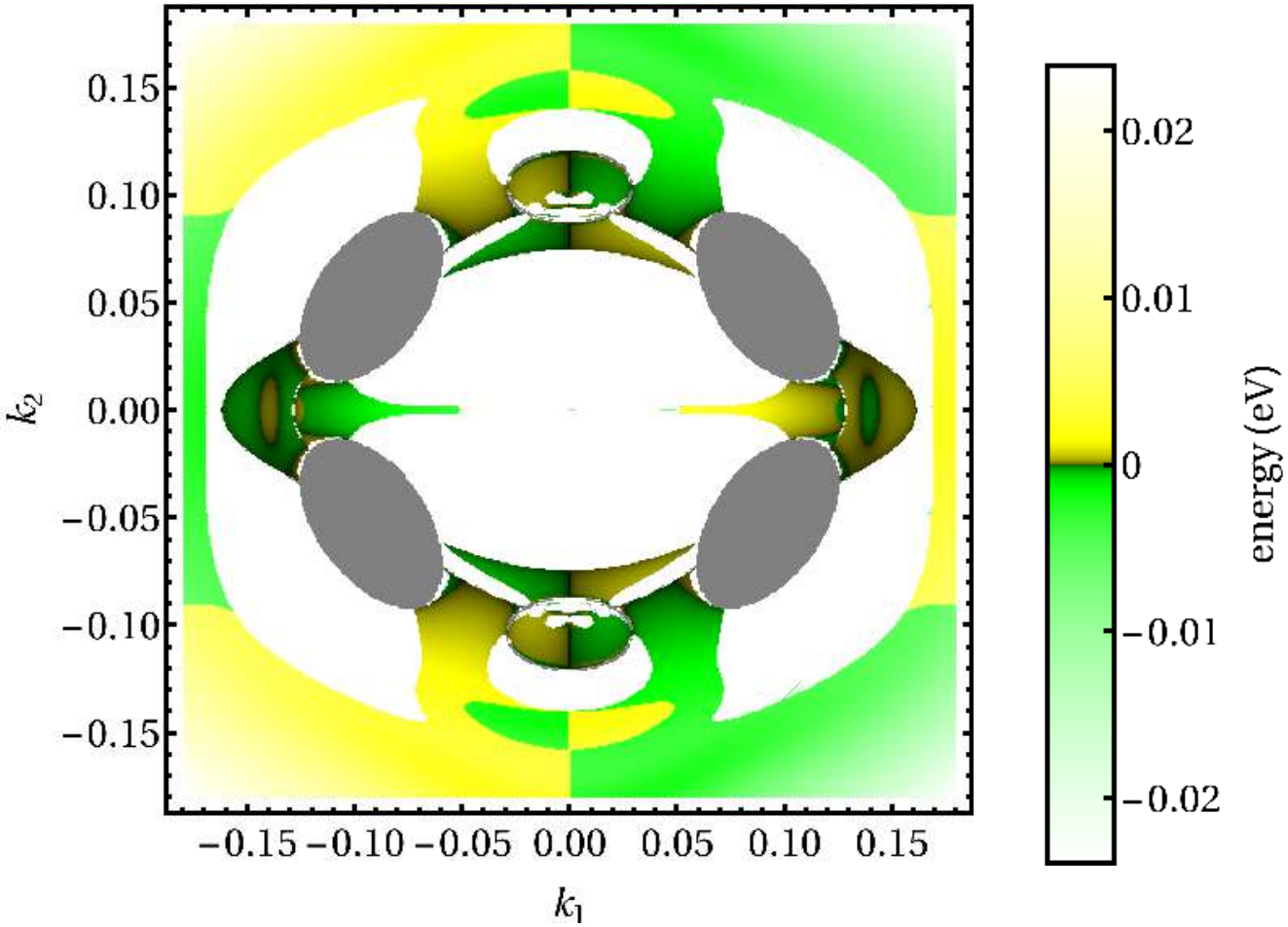}\\[2ex]%
\raisebox{1ex}{(b)}\includegraphics[width=0.95\columnwidth]{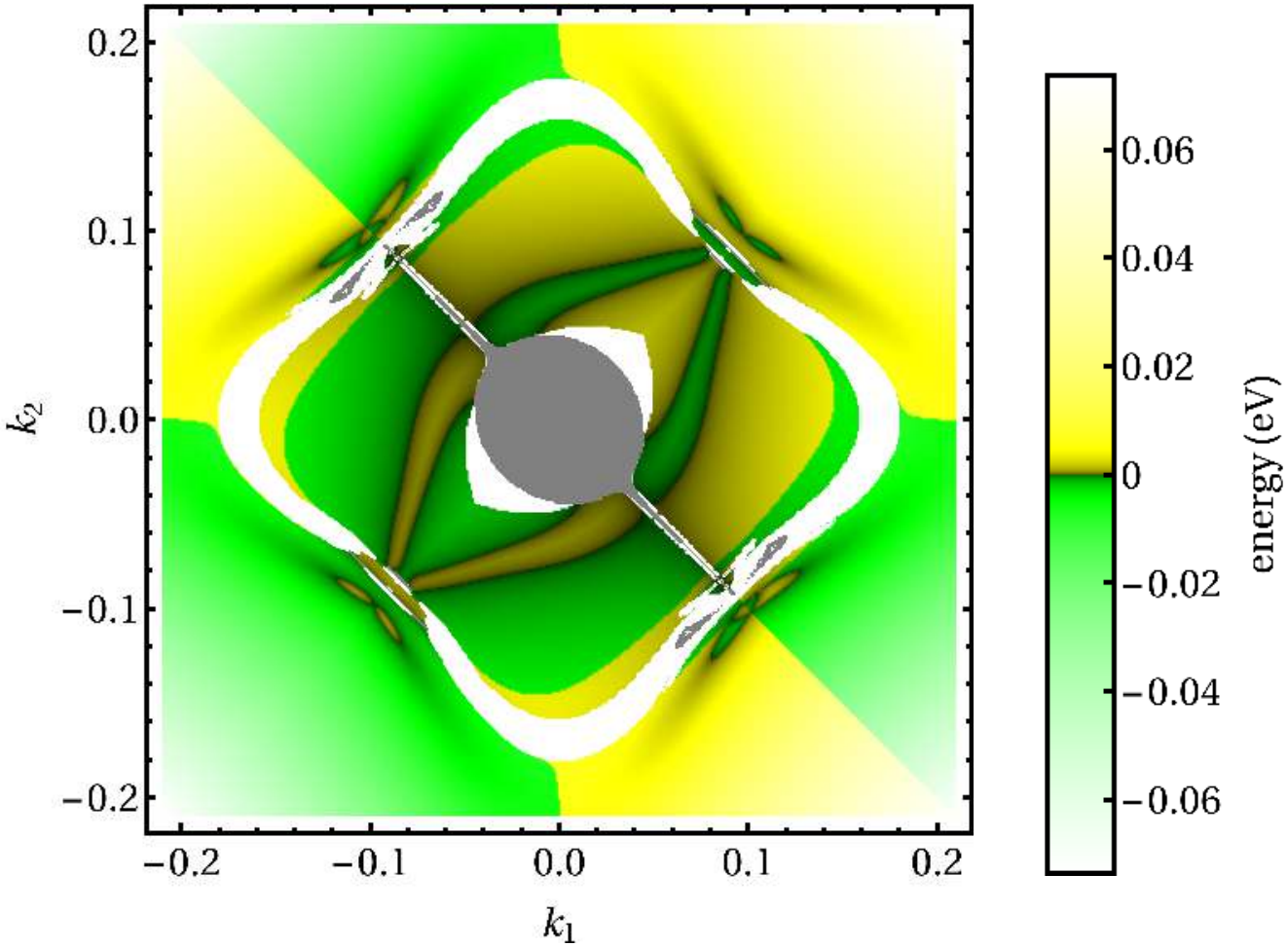}
\end{center}
\caption{Dispersion of surface states of YPtBi in the superconducting $T_2$ state for (a) the (111) surface with a thickness of $W=16\,000$ and (b) the (100) surface with $W=4000$. A larger thickness was considered for the (111) surface because of large finite-size effects. The spectra at fixed momentum are not symmetric. The signful energy closest to the Fermi energy is plotted. The gray regions are the projections of the Bogoliubov Fermi pockets shown in Fig.\ \ref{fig.T2.nodes} onto the surface Brillouin zone. In these regions, states exist at the Fermi energy and are bulk like. In the white regions, there are no Fermi pockets but the states closest to the Fermi energy are still bulk like.}
\label{fig.T2.surface}
\end{figure}

\subsubsection{Surface states}

Figure \ref{fig.T2.surface} shows the dispersion of surface states at
the (111) and (100) surfaces. Note that for the $T_2$ pairing state,
(100) is not equivalent to (001). Outside of the projection of the
normal-state Fermi surface, we again find weakly modified descendants
of the normal surface states shown in
Fig.\ \ref{fig.normal.surface}. Also visible are the projections of
the inflated nodes allowed by the broken time-reversal symmetry. There
are no flat bands associated with the line nodes in the high-symmetry
plane. Flat bands are not expected since the line nodes are not protected
by a winding number~\cite{KTS15}, unlike the line nodes in the
$A_1$ state, but by a
$\mathbb{Z}_2$ invariant that only exists in the $k_xk_y$
plane. Hence, this $\mathbb{Z}_2$ number does not induce a global
invariant on gapped one-dimensional subsystems and one cannot
construct an argument for flat bands in analogy to the $A_1$ case. 
There are also no flat bands from the inflated nodal rings that are not lying in the symmetry plane, i.e., the small orange rings at the top and bottom of Fig.\ \ref{fig.T2.nodes}(a). Their projections are visible in Fig.\ \ref{fig.T2.surface}(a) as two ellipses that cross the $k_1=0$ line and intersect the projections of the outer
nodal line in the symmetry plane. We conclude that these inflated
rings are also not protected by a winding number. 

Furthermore, we observe dispersive surface states where the
normal-state Fermi sea has been gapped out. Their dispersion crosses
the Fermi energy at arclike lines emanating from the projections of
the point nodes for both the (111) and the (100) slab.
These arcs show up as black lines separating yellow and green regions in Figs.\ \ref{fig.T2.surface}(a) and \ref{fig.T2.surface}(b).
For the (111) slab, Fig.\ \ref{fig.T2.surface}(a), the projections of the point nodes lie within the projections of the inflated rings. Two arcs start from each point node, consistent with its Chern number $\pm 2$. One arc, which is clearly visible, connects to the inflated ring. The other connects to the projection of the outer ring in the $k_xk_y$ plane and is obscured by bulk states (white region).
The arcs are straight lines lying in the projection of a mirror plane,
cf.\ Sec.\ \ref{sec.A1}. Due to the mirror symmetry, the spectrum and
in particular the arcs are twofold degenerate. Correcting for the double counting in the Nambu formalism, this corresponds to a single arc; a single pair of helical Majorana bands crosses here.
For the (100) slab, Fig.\ \ref{fig.T2.surface}(b), four arcs are
associated with each point node, where the point nodes themselves are
obscured by bulk states. The appearance of four arcs instead of
two can be understood as follows: The centrosymmetric variant of the
model has a symmetric spectrum at each $\mathbf{k}_\|$ and thus
wherever a Majorana surface band with dispersion $E(\mathbf{k}_\|)$ crosses the
Fermi energy, forming an arc, another band with dispersion
$-E(\mathbf{k}_\|)$ also crosses the Fermi energy. In the present case
the centrosymmetric variant has two arcs associated with each point
node. The simultaneous breaking of inversion and time-reversal
symmetry shifts this crossing to finite energy and thus splits each
arc into two. This is consistent with the observation that neighboring
arcs in Fig.\ \ref{fig.T2.surface}(b) have opposite velocities. The
crossing of the surface bands happens between these arcs.

\section{Summary and conclusions}
\label{sec.concl}

Bulk and surface states of two plausible superconducting states of
half-Heusler compounds have been analyzed, taking YPtBi as a specific
example. Their partially filled $\Gamma_8$ band is described in terms
of the effective angular momentum $j=3/2$. The inverted band structure
of YPtBi and several other half-Heusler compounds
\cite{CQK10,LWX10,LYW16} leads to the appearance of surface states of
topological origin even in the normal phase. These compounds are also
noncentrosymmetric and thus allow us to study the fate of
characteristic properties of noncentrosymmetric superconductors, such
as flat surface bands, in a multiband system. YPtBi is a particularly
promising candidate for topological superconductivity based on
experimental reports of a zero-bias peak in tunneling \cite{KWN16},
which hints at such flat surface bands. Technically, we have performed
numerical diagonalization of Bogoliubov-de Gennes Hamiltonians for
slabs with (111) and (100) oriented surfaces. 

We consider a $A_1$ pairing state that leaves time-reversal symmetry
intact and a $T_2$ pairing state that breaks time-reversal and also
lattice symmetries. The two states have in common that they allow for
line nodes of the superconducting gap, which are supported by
measurements of the London penetration depth \cite{KWN16}. 

For the $A_1$ state, the line nodes require the symmetry-allowed
admixture of nonlocal \textit{p}-wave pairing \cite{BWW16}. The $A_1$
state has six nodal rings surrounding the cubic coordinate axes. The
nodal rings are associated with nonzero winding numbers $\pm 1$. These
protect flat zero-energy surface bands bounded by the projections of a
\emph{single} nodal ring onto the surface Brillouin zone. Such flat
bands have been found for single-band noncentrosymmetric models
\cite{BST11,ScR11,SBT12,ScB15,BTS13,STB13}. We here find that they
 persist in the multiband case, in particular, they are not gapped out
by interband pairing. On the other hand, the multiband character
allows for additional pairing states that are not possible for
single-band superconductors. The $A_1$ state considered here is a
superposition of singlet ($J=0$) and septet ($J=3$) pairing, where the
latter is possible because of the effective spin $j=3/2$ of the
electrons. In addition, we have obtained dispersive surface
states. They are in part derived from the normal-phase surface states,
but interesting new effects emerge where the normal-phase Fermi sea is
gapped out by superconductivity. Here, Fermi arcs appear when mirror
planes are perpendicular to the surface. They are restricted to the
projection of the mirror plane onto the surface Brillouin zone and are
protected by a mirror parity. 

The time-reversal-symmetry-breaking $T_2$ state has interesting nodal
structure already in the bulk. Even for infinitesimal pairing, the
ASOC changes the nodal structure compared to the centrosymmetric
variant of the model or the case of vanishing ASOC \cite{BWW16}:
besides point nodes generic for topological superconductors in class D
and line nodes in a high-symmetry plane, which rely on a twofold
rotation axis, the system has additional nodal rings, reminiscent of
the $A_1$ case. If the pairing amplitude is not infinitesimal we find
that the point nodes and the line nodes in the high-symmetry plane
survive but now coexist with two-dimensional Bogoliubov Fermi pockets
(inflated nodes). At
surfaces, the $T_2$ superconductor does not show flat bands, due to
the lack of winding numbers that could protect them. There are
dispersive surface states with Fermi arcs associated with projections
of the point nodes. The arcs for the (100) surface are split due to
the absence of  time-reversal and inversion symmetry so that their
number is doubled compared to systems with time-reversal or inversion
symmetry and the same Chern numbers of the point nodes. For the (111)
surface, the arcs lie in a mirror plane, which prevents the
splitting. 

\textit{Note added}: Recently, a preprint by Yang \textit{et al.}\ \cite{Yang} appeared that also addresses the flat-band surface states for $A_1$ pairing.

\begin{acknowledgments}

The authors thank J. Paglione, H. Kim, and H. Nakamura for stimulating discussions.
C.~T. acknowledges financial support by the Deutsche Forschungsgemeinschaft, in part through Research Training Group GRK 1621 and Collaborative Research Center SFB 1143.
D.~A. acknowledges support from the National Science Foundation Grant No.\ DMREF-1335215 and the hospitality of the Pauli Center of the ETH Z{\"u}rich.
P.~M.~R.~B. acknowledges the hospitality of the TU Dresden.

\end{acknowledgments}

\appendix

\section{Expansion of the determinant of the Bogolibov-de Gennes Hamiltonian}
\label{app.det}

We here show that the coefficient
\begin{equation}
g_2 = \frac{1}{2}\,\frac{d^2}{d\Delta^2}\, \det \mathcal{H} \bigg|_{\Delta=0}
\end{equation}
in the expansion (\ref{T2.detexp.2}) is nonnegative everywhere on the Fermi surface and is zero where the cubic axes intersect the Fermi surface. The short-hand notation $\Delta=\Delta^0_{T_2}$ is used and the momentum argument is suppressed.

Take $E$ to be a real energy but not an eigenvalue of $\mathcal{H}(\mathbf{k})$. Then $\mathcal{H}(\mathbf{k})-E$ is invertable (an identity matrix is suppressed) and we have
\begin{equation}
\frac{d}{d\Delta}\, \det (\mathcal{H}-E) = \det (\mathcal{H}-E)
  \Tr (\mathcal{H}-E)^{-1}\, \frac{d\mathcal{H}}{d\Delta}
\end{equation}
and
\begin{align}
\frac{d^2}{d\Delta^2}\,& \det (\mathcal{H}-E) = \det (\mathcal{H}-E)
  \left[ \Tr (\mathcal{H}-E)^{-1}\, \frac{d\mathcal{H}}{d\Delta} \right]^2  \nonumber \\
&{}- \det (\mathcal{H}-E) \Tr (\mathcal{H}-E)^{-1}\, \frac{d\mathcal{H}}{d\Delta}
    (\mathcal{H}-E)^{-1}\, \frac{d\mathcal{H}}{d\Delta} .
\end{align}
Note that
\begin{equation}
\frac{d\mathcal{H}}{d\Delta} = 4\left(\begin{array}{cccc|cccc}
  & & & & 0 & 0 & 0 & 0 \\
  & & & & 0 & 0 & 0 & i \\
  \multicolumn{4}{c|}{\raisebox{1.5ex}[-1.5ex]{0}} & 0 & 0 & 0 & 0 \\
  & & & & 0 & -i & 0 & 0 \\ \hline
  0 & 0 & 0 & 0 & & & & \\
  0 & 0 & 0 & i & & & & \\
  0 & 0 & 0 & 0 & \multicolumn{4}{c}{\raisebox{1.5ex}[-1.5ex]{0}} \\
  0 & -i & 0 & 0 & & & &
  \end{array}\right) .
\end{equation}
Setting $\Delta=0$ we thus get
\begin{align}
\frac{d}{d\Delta}\,& \det (\mathcal{H}-E)\bigg|_{\Delta=0}
  = \Tr \left(\begin{array}{cc}
    h - E & 0 \\ 0 & -h^T - E
  \end{array}\right)^{\!-1} \frac{d\mathcal{H}}{d\Delta} \nonumber \\
&= \Tr \left(\begin{array}{cc}
    (h - E)^{-1} & 0 \\ 0 & (-h^T - E)^{-1}
  \end{array}\right) \frac{d\mathcal{H}}{d\Delta} = 0
\end{align}
and, using this,
\begin{align}
\frac{d^2}{d\Delta^2}\,& \det (\mathcal{H}-E)\bigg|_{\Delta=0}
  = - \det (\mathcal{H}-E) \nonumber \\
&{}\times \Tr (\mathcal{H}-E)^{-1}\, \frac{d\mathcal{H}}{d\Delta}
    (\mathcal{H}-E)^{-1}\, \frac{d\mathcal{H}}{d\Delta}\bigg|_{\Delta=0} .
\end{align}
We denote the eigenvalues and eigenvectors of $\mathcal{H}$ for $\Delta=0$ by $E_i$ and $|i\rangle$, respectively. The spectrum is symmetric; we enumerate the eigenvalues in such a way that $E_{-i}=-E_i$. Then
\begin{align}
\frac{d^2}{d\Delta^2}\,& \det (\mathcal{H}-E)\bigg|_{\Delta=0}
  = - \prod_k (E_k-E) \nonumber \\
&{}\times \sum_{ij} \frac{\big|\langle i|d\mathcal{H}/d\Delta|j\rangle\big|^2}
    {(E_i-E)(E_j-E)} .
\label{appb.g2gen.3}
\end{align}
Now take $\mathbf{k}$ on the normal-state Fermi surface. There are two cases: The corresponding normal-state eigenvalue can be nondegenerate or twofold degenerate. The latter happens at the intersection of the Fermi surface with the cubic axes.

Case 1: nondegenerate eigenvalue. We take $E_1=E_{-1}=0$ and write
\begin{equation}
\prod_k (E_k-E) = E^2 \prod_{k\neq \pm 1} (E_k-E) .
\end{equation}
In the sum over $i$, $j$, the terms with $i,j\in \{1,-1\}$ have the denominator $E^2$, which cancels with the $E^2$ in the prefactor. The terms with only one of $i$ or $j$ from $\{1,-1\}$ contain only one factor $E$ in the denominator, with leaves an overall factor of $E$. The terms with $i,j\notin\{1,-1\}$ retain a prefactor of $E^2$. We now take the limit $E\to 0$, i.e., $E$ goes to the Fermi energy. Only the first term survives so that
\begin{equation}
g_2 = \frac{1}{2}\, \frac{d^2}{d\Delta^2}\, \det \mathcal{H} \bigg|_{\Delta=0}
 \! = - \frac{1}{2} \prod_{k\neq \pm 1} E_k \sum_{i,j= \pm 1}
    \left|\langle i| \frac{d\mathcal{H}}{d\Delta} |j\rangle\right|^2 .
\end{equation}
Using that the spectrum is symmetric, this yields
\begin{equation}
g_2 = \frac{1}{2}\, E_2^2 E_3^2 E_4^2 \sum_{i,j= \pm 1}
  \left|\langle i| \frac{d\mathcal{H}}{d\Delta} |j\rangle\right|^2
  \ge 0 .
\end{equation}  
  
Case 2: degenerate eigenvalue. We take $E_{\pm 1}=E_{\pm 2}=0$ and write
\begin{equation}
\prod_k (E_k-E) = E^4 \prod_{k\neq \pm 1,\pm 2} (E_k-E) .
\end{equation}
The terms in the sum over $i$, $j$ in Eq.\ (\ref{appb.g2gen.3}) are at most of order $1/E^2$ for small $E$. Hence, all terms vanish in the limit $E\to 0$ and we obtain $g_2=0$.


\begin{thebibliography}{99}
\bibitem{endnote.Chern}The Chern number $\pm 2$ is obtained by taking the Bo\-go\-liu\-bov-de Gennes Hamiltonian literally, which agrees with the convention used in Ref.\ \cite{BzS17}. A different convention is used in Ref.\ \cite{ScB15}, which corresponds to dividing the Chern number by $2$ to account for the double counting of degrees of freedom in the Nambu formalism. These different conventions also lead to conflicting meanings of the term ``double Weyl point'' in these papers.
\end{thebibliography}
\end{document}